\pdfoutput=1

\documentclass[11pt,twoside,a4paper,cmspaper,final,collab]{cms-tdr}
\def\svnVersion{183693}\def\cmsCernNo{2012-341}\def\cmsCernDate{\today}\def\cmsMessage{Submitted to Physical Review D}

\begin{document}\cmsNoteHeader{SUS-11-018}

\hyphenation{had-ron-i-za-tion}
\hyphenation{cal-or-i-me-ter}
\hyphenation{de-vices}

\RCS$Revision: 174260 $
\RCS$HeadURL: svn+ssh://svn.cern.ch/reps/tdr2/papers/SUS-11-018/trunk/SUS-11-018.tex $
\RCS$Id: SUS-11-018.tex 174260 2013-02-27 15:01:25Z nsaoulid $
\providecommand{\mt}{\ensuremath{M_\cmsSymbolFace{T}}\xspace}
\ifthenelse{\boolean{cms@external}}{\providecommand{\cmsLeft}{Top}\providecommand{\cmsRight}{Bottom}}{\providecommand{\cmsLeft}{Left}\providecommand{\cmsRight}{Right}}
\newlength\cmsFigWidth
\ifthenelse{\boolean{cms@external}}{\setlength\cmsFigWidth{\columnwidth}}{\setlength\cmsFigWidth{0.77\textwidth}}
\cmsNoteHeader{SUS-11-018} 
\title{Search for supersymmetry in events with opposite-sign dileptons and missing transverse energy using an artificial neural network}

\author[IC] {Michele Pioppi}

\date{\today}

\abstract{
In this paper, a search for supersymmetry (SUSY) is presented in events
with two opposite-sign isolated leptons in the final state, accompanied
by  hadronic jets and missing transverse energy. An artificial neural
network is
employed to discriminate possible SUSY signals from
standard model background. The analysis uses a data sample collected
with the CMS detector during the 2011 LHC run, corresponding to an
integrated luminosity of  4.98\fbinv of proton-proton collisions at the
center of mass energy of 7\TeV.
Compared to other
 CMS analyses, this one uses relaxed criteria on missing transverse energy
($\ETslash >40\GeV$) and total hadronic transverse energy ($\HT>120\GeV$), thus
 probing
 different regions of parameter space. Agreement is found between
 standard model expectation and observations, yielding limits in the context
 of the constrained minimal supersymmetric standard model and on a set of
 simplified models.
}

\hypersetup{%
pdfauthor={CMS Collaboration},%
pdftitle={Search for supersymmetry in events with opposite-sign dileptons and missing transverse energy using an artificial neural network},%
pdfsubject={CMS},%
pdfkeywords={CMS, physics, supersymmetry, neural network}}

\maketitle 

\section{Introduction}
\label{chap:Intro}

 One of the most natural extensions of the standard model (SM) of particle physics is
 supersymmetry (SUSY)~\cite{ref:SUSY-1, ref:SUSY0, ref:SUSY1, ref:SUSY2, ref:SUSY3,
 ref:SUSY4, ref:hierarchy1, ref:hierarchy2}. Supersymmetry allows for gauge coupling unification at the energy of  $10^{16}\GeV$,  provides a good dark matter candidate (lightest  supersymmetric particle, LSP)~\cite{ref:SUSY-LSP-DM},  is a
 necessary component to explain quantum gravity in the framework of string theory,
 and automatically cancels the quadratic divergences in radiative corrections to the
 Higgs boson mass.
For every particle in the
 standard model, SUSY
  introduces a super-partner, the ``sparticle", with spin differing by $1/2$
unit from the SM particle.
 There are theoretical arguments that suggest sparticle
 masses could be less than ${\sim}1$\TeV~\cite{ref:hierarchy1,
 ref:hierarchy2} making the experiments at the
 Large Hadron Collider (LHC) an ideal place
 for their discovery.

 With the successful 2011 LHC run, an integrated luminosity of 4.98\fbinv in $\Pp\Pp$ collisions
 at 7\TeV center-of-mass energy has been collected with the Compact Muon Solenoid (CMS)
 experiment.  This dataset is used to search for the presence of SUSY particles
 in
 events with two opposite-sign leptons (electrons and muons) in the final state,
 utilizing an artificial neural network (ANN).
 Two opposite-sign leptons can be produced  in a SUSY cascade through the decay
 of neutralinos and charginos.
 Assuming that
 $R$-parity is conserved~\cite{Farrar:1978xj}, a stable, weakly
 interacting LSP exists, resulting in a missing transverse energy (\ETslash ) signature.
 The amount of missing transverse energy depends on the mass splittings among the heavier
 sparticles.
 So far, typical dilepton SUSY searches in CMS  have required several
 jets with large transverse momentum, which correspond to large values of \HT,
 the scalar  sum over the transverse momenta of all jets satisfying the jet selection,
and large missing transverse energy
to discriminate a SUSY signal from the very large SM backgrounds.
Compared with previous CMS searches~\cite{ref:ref_011,ref:ref_012}, this analysis uses relaxed
criteria on missing transverse energy ($\ETslash > 40\GeV$) and \HT
($\HT > 120\GeV$).
For SUSY models that yield events with large
\ETslash, the ANN's performance is comparable to the data analyses
using large \ETslash and \HT. Hence, for such models the
additional power of a multivariate technique is not required to discriminate
between new
physics and the SM backgrounds.  However, for SUSY models that yield low-\ETslash or low-\HT signatures, the
discriminating power of the ANN helps to suppress the large SM backgrounds.

 The results are interpreted in the context of  the constrained minimal
 supersymmetric  standard model (CMSSM
~\cite{ref:CMSSM1,ref:CMSSM2}), and
a class of simplified model scenarios (SMS)~\cite{ref:SMS2,ref:SMS1}. For
illustration purposes, the
benchmark CMSSM point LM6 ($m_0 = 85$\GeV, $m_{1/2} = 400$\GeV, $\tan\beta=10$, $A_0 =
0$\GeV) is used throughout the paper. In the class of SMS considered, gluinos are pair-produced, with one
of them decaying as $\PSg \rightarrow \PSGczDt j j
\rightarrow \PSGczDo
\ell^{+}\ell^{-} j j $,
and the other as~$\PSg \rightarrow \PSGczDo j j$. Here
$\PSGczDt$ is the second-lightest neutralino, $\PSGczDo$ is the lightest
neutralino and the LSP, and $\ell = \Pe$, $\mu$, or $\tau$ with equal
probability. This SMS thus
always leads to a pair of opposite-sign leptons in the final state, in addition to the
jets and \ETslash. The SMS is fully described by the following parameters: the
masses of the gluino ($m_{\PSg}$), and the LSP ($m_\text{LSP}$), along with the
neutralino mass in the gluino decay which is set to
$m_{\PSGczDt}=(m_{\PSg}+m_\text{LSP})/2$.

\section{CMS Detector}

 A detailed description of the CMS Detector can be found elsewhere
~\cite{ref:ref_013b}. A  right-handed coordinate system is used with the
 origin at the nominal interaction point. The $x$ axis points to the
 center of the LHC  ring, the $y$ axis is vertical and points upward,
 and the $z$ axis points in the direction of the
counterclockwise proton beam. The
 azimuthal angle  $\phi$ is measured
  with respect to the $x$ axis in
 the $x$-$y$ plane and the polar angle $\theta$ is defined with respect to the
 $z$ axis, while the  pseudorapidity is defined as $\eta = - \ln[\tan
 (\theta/2)]$. The central feature of the CMS apparatus is a
 superconducting solenoid, of  6\unit{m}  internal diameter, that produces a
 magnetic   field of 3.8\unit{T}. Located within the field volume are the silicon pixel
 and strip tracker, and the  barrel and endcap calorimeters ($|\eta| < 3$),
 composed of a crystal electromagnetic calorimeter (ECAL) and a brass and
 scintillator hadron calorimeter (HCAL). Calorimetry provides
 energy and direction measurements of electrons and hadronic
 jets.  The detector is nearly  hermetic, allowing for energy balance
 measurements in the plane transverse to the beam directions. Outside the
 field volume, in the forward region ($3 < |\eta| < 5$), there is
 an iron and quartz-fiber hadron calorimeter. The  steel return yoke outside
 the solenoid is instrumented with gas-ionization detectors used to
 identify muons. The CMS experiment collects data using a two-level trigger system, the
 Level-1 (L1) hardware trigger~\cite{ref:ref_014} and a high-level
 software trigger (HLT)~\cite{ref:ref_015}.
\section{Data Samples, Trigger, and Event  Selection}

Data events are selected
using a set of  dilepton  triggers, which require
the presence of at least two leptons, either two muons or two electrons or a
muon-electron pair. In the case of the double-muon trigger, the selection is
asymmetric  with a transverse momentum (\pt) threshold
of 13\GeV for the leading (higher-\pt) muon and 8\GeV for the subleading
one. In the case of the double-electron trigger, the selection is asymmetric with a
threshold
applied to the transverse energy of a cluster in the ECAL.
The thresholds are fixed to 17\GeV (8\GeV) for the leading (subleading) electron energy.
For the muon-electron trigger, the threshold on the transverse momentum,
\pt (transverse energy, \et) is 8\GeV (17\GeV)  for the muon (electron).
For all triggers, additional identification
and isolation
criteria are also applied.

Muon candidates are reconstructed \cite{ref:Muon} by combining the information from the inner tracking system, the calorimeters, and the muon system.
Electron candidates are reconstructed \cite{ref:Electron} by combining the information from the ECAL with the silicon tracker, using shower shape and
track-ECAL-cluster matching variables in order to increase the sample
purity. Jets are reconstructed using the anti-\kt clustering algorithm
~\cite{ref:ref_016} with a distance parameter $\Delta R=\sqrt{(\Delta \phi)^2 + (\Delta \eta)^2 }=0.5$.
The inputs to the jet clustering  algorithm are the four-momentum vectors
of reconstructed particles. Each such particle is reconstructed with the
particle-flow  technique~\cite{ref:PFLOW} which combines information from several subdetectors.
The measured jet transverse
momenta are corrected with scale factors derived from
simulation; to correct  for any differences in the energy response
between simulation and data, a residual correction factor derived from the latter is
applied to jets in the data~\cite{ref:ref_017}.
In general, $\ETslash \equiv -\abs{\sum \ptvec}$, where the sum is
taken over all final-state particles reconstructed in the
CMS detector.
The total transverse energy ($\sum \ET$) of the event is
calculated as the scalar sum of the transverse energies of leptons and jets.
The total hadronic transverse energy, ($\HT\equiv |\sum
\ptvec|$), is computed as the  scalar sum of the transverse
energies of all reconstructed jets in the  event satisfying the jet
selection criteria described below.

Simulated $\Pp\Pp$ collision events are produced with the {\PYTHIA}
6.4.22~\cite{ref:ref_018} generator (using underlying event tune Z2
which is identical to the Z1 tune~\cite{D6T} except that Z2
uses the  CTEQ6L parton distribution functions (PDF)  while Z1 uses CTEQ5L) for QCD, $\PW\PW$, $\cPZ\cPZ$ and $\PW\cPZ$ samples.
For \ttbar, Drell--Yan, and  $\PW+\text{jets}$ samples  the {\MADGRAPH 4.4.24}~\cite{ref:ref_019} generator is used.
Events are then processed with a simulation of the CMS detector response based on {\GEANTfour}~\cite{ref:ref_020}. Multiple proton-proton interactions are superimposed
on the hard collision, and all simulated event samples are reweighted according
to the distribution of the number of
reconstructed primary vertices in data. Simulated events are reconstructed and analyzed
in the same way as data events. Simulated event samples are
used to train the ANN, to extrapolate background
estimates from a background-enriched control region in data to the expected signal-enriched region, and
to estimate systematic uncertainties.

Non-collision backgrounds are removed by applying  quality requirements ensuring the presence of at least one reconstructed primary
vertex~\cite{ref:ref_021}.
Events are required to have at least two opposite-sign leptons, both electrons or muons, or an electron-muon pair, with $\pt 
> 20\GeV$ and
$|\eta| < 2.4$, and at least two jets with $\pt > 30$\GeV and
$|\eta| < 2.4$.  Jets are required to satisfy  the quality criteria
described in Ref.~\cite{ref:refcalo}. Leptons are
required to be isolated from significant energy deposits and tracks in a
cone of radius $\Delta R = 0.3$ around the direction of the lepton. The
relative combined isolation, defined as $I_\text{rel}^\text{comb} =
(\sum_\text{tracks}
\pt+ \sum_\text{ECAL} \ET + \sum_\text{HCAL}
\ET)/\pt$, is required to be
$<$0.2 for muons and $<$0.08 for electrons, with the latter criterion being
more strict in order to reject jets misidentified as electrons.
\section{Signal to Background Discrimination}

The ANN in this analysis is used to separate SUSY signals from  SM background events,
exploiting correlations among the discriminating variables, and thus
providing
improved results with respect to the use of sequential selections.
Due to the presence of isolated leptons, the main SM
background contributions to this analysis involve the
production of \ttbar, and $\cPZ+\text{jets}$. The QCD multijet processes with two
misidentified (fake) leptons, and $\PW+\text{jets}$ events with one misidentified lepton
can also be part of the background, but are significantly reduced by applying additional
candidate event selection
criteria described below. Finally, two leptons in the final state could be produced by  $\PW\PW$, $\PW\cPZ$ or $\cPZ\cPZ$ decays
but their  contributions are found in simulation to be negligible compared to the main backgrounds.

The candidate event selection criteria, which are imposed before the ANN training, are the following: events are required
to have  $\ETslash > 30$\GeV, the distance $\Delta R$
between either of the two leading opposite-sign leptons and the
closest jet is required to be $>$0.2, and
the dilepton mass $M_{\ell\ell}$, formed from the two leading
opposite-sign leptons, is required to be larger than 10\GeV.
These criteria reject  the vast majority of the background, while retaining most of the  signal as shown
in Table
\ref{ref:table3} for
CMSSM benchmark point LM6.
This greatly facilitates the ANN training and optimization by
excluding a region  heavily  dominated by background in which
few if any signal events are present. The signal region is defined by the candidate event selection
criteria with  an additional requirement on the ratio of the dilepton transverse
energy $\sum \ET^\text{lepton}$ to the total transverse energy (as defined in
Section 3) to be less than 0.4.

\begin{table}[htbH]
\begin{center}
\footnotesize
\topcaption{Expected number of signal and background (bkg.) events after the event selection
criteria, and after the candidate event selection criteria for events in the
signal region are applied.
 The next-leading-order (NLO) cross section is used for the CMSSM benchmark point LM6 yield
determination. The dataset resulting from the candidate event selection is
used as input to the ANN. The uncertainties quoted are statistical only.}
\begin{scotch}{cD{,}{\,\pm\,}{6.4}D{,}{\,\pm\,}{5.3}}
 Sample & \multicolumn{1}{c}{Event Selection} & \multicolumn{1}{c}{Signal Region} \\
 \hline
 \ttbar      &   17395,60    &  8271,40  \\
 $\cPZ+\text{jets}$        &  507316,1200  &  4740,60  \\
 $\PW+\text{jets}$        &   21094,740   &   416,40  \\
 $\PW\PW$              &    1204,10    &    15,1   \\
 $\PW\cPZ$              &    1750,8     &    20,1         \\
 $\cPZ\cPZ$              &    1225,4     &    13,1         \\
 QCD             &   19578,7500  &  1313,260 \\
 Total SM Bkg.   &  569562,7700  & 14797,280 \\
 LM6             &      71,1     &    54,1   \\
\end{scotch}
\label{ref:table3}
\end{center}
\end{table}

The ANN training samples are based on simulated events.
A mixture of \ttbar, $\cPZ+\text{jets}$, $\PW+\text{jets}$, and QCD simulated samples
are used as the SM background. For the signal,  a class of SMS scenarios~\cite{ref:SMS2} is used.
For the ANN training grid points close to the diagonal ($m_{\PSg}$ = $m_\text{LSP}$)
are used  with $|m_{\PSg}-m_\text{LSP}|<400\GeV$. These points  are
chosen since they exhibit  low \ETslash or \HT thresholds: more than
90\% of the events have $\ETslash <200\GeV$ or $\HT <600\GeV.$

Several topological and kinematical variables are considered
according to their potential to discriminate SM backgrounds from possible
SUSY signals, taking into account the correlations among them. The variables studied
are based on the general production and decay characteristics of many
supersymmetric processes and are not tuned to a specific model.

Using different combinations of candidate input variables,
several ANNs are constructed and compared in order to select
the optimal configuration.  The differences in
performance are studied and quantified in terms of the signal selection efficiency as a function of
background rejection.
A network with seven input
variables, those with the smallest degree of
correlation among themselves and with the
highest discriminating power, shows the best performance.
The ANN variable importance is defined as sum of the weights-squared of the connections between the
variable's neuron in the input layer and the ones in the first hidden layer.
Table \ref{tableAdd} lists the seven input ANN variables along with their
description, and their relative importance after the ANN training.

\begin{table*}[htbH]
\begin{center}
\footnotesize
\topcaption{Seven event, lepton and jet related variables used for the ANN
construction.The transverse mass $\mt$ is defined as $\sqrt{(\sum \ET)^2 -
(\sum \ptvec)^2}$, where $\sum \ET$ and $\sum \ptvec$ represent the scalar and
vector sums over the transverse momenta of all reconstructed jets and
leptons.}
\begin{scotch}{ccc}
 Variable & Description  & ANN weight (\%)\\ \hline
 \ETslash                          & Missing transverse energy               & 22  \\
 $M_{\ell\ell}$                          & Dilepton mass                            & 20 \\
 $\frac{\sum{\ET^\text{lepton}}}{\sum{\ET}}$  & Ratio of the energy of the dilepton system to total transverse energy & 18 \\
 $N_\text{jets}$                        & Number of jets                          & 13 \\
 Jet2 $\pt$                        & Subleading jet $\pt$                     & 12 \\
 $\mt$                             & Transverse mass                    &  8 \\
 Jet1 $\pt$                        & Leading jet $\pt$                       &  7 \\
\end{scotch}
\label{tableAdd}
\end{center}
\end{table*}

\section{ANN Output for SM Background }
\label{SMTemplate}

 In order to quantify the level of agreement and the significance of a possible excess
between data and SM expectation, it is important to provide a robust estimate of the ANN output
distribution  in the signal region under the SM-only hypothesis along with its systematic uncertainty.

The approach used to estimate the ANN  prediction for the SM-only hypothesis from data is as follows.
A signal region (SR) is defined by the set of the candidate event selection requirements
and the additional criterion on the fraction of transverse energy carried by the dilepton system as described
in Section 3. A primary control region (CR) is defined by inverting two of the signal event selection
criteria, the total missing transverse energy and the selection cut on the
fraction of transverse energy carried by the dilepton system. This region is chosen
so that it is dominated by SM processes. Signal contamination in the primary control
region is small: for the LM6 benchmark point it is less than 0.03\%, and less than 0.4\% for SMS points
close to the diagonal ($m_{\PSg} = m_\text{LSP}$).
The ANN output distribution in the  primary control region is then
obtained using data $\mathrm{ANN(SM)}^\text{data}_\text{CR}$.

\begin{figure}[ht]
\centering
\includegraphics[width=0.48\textwidth]{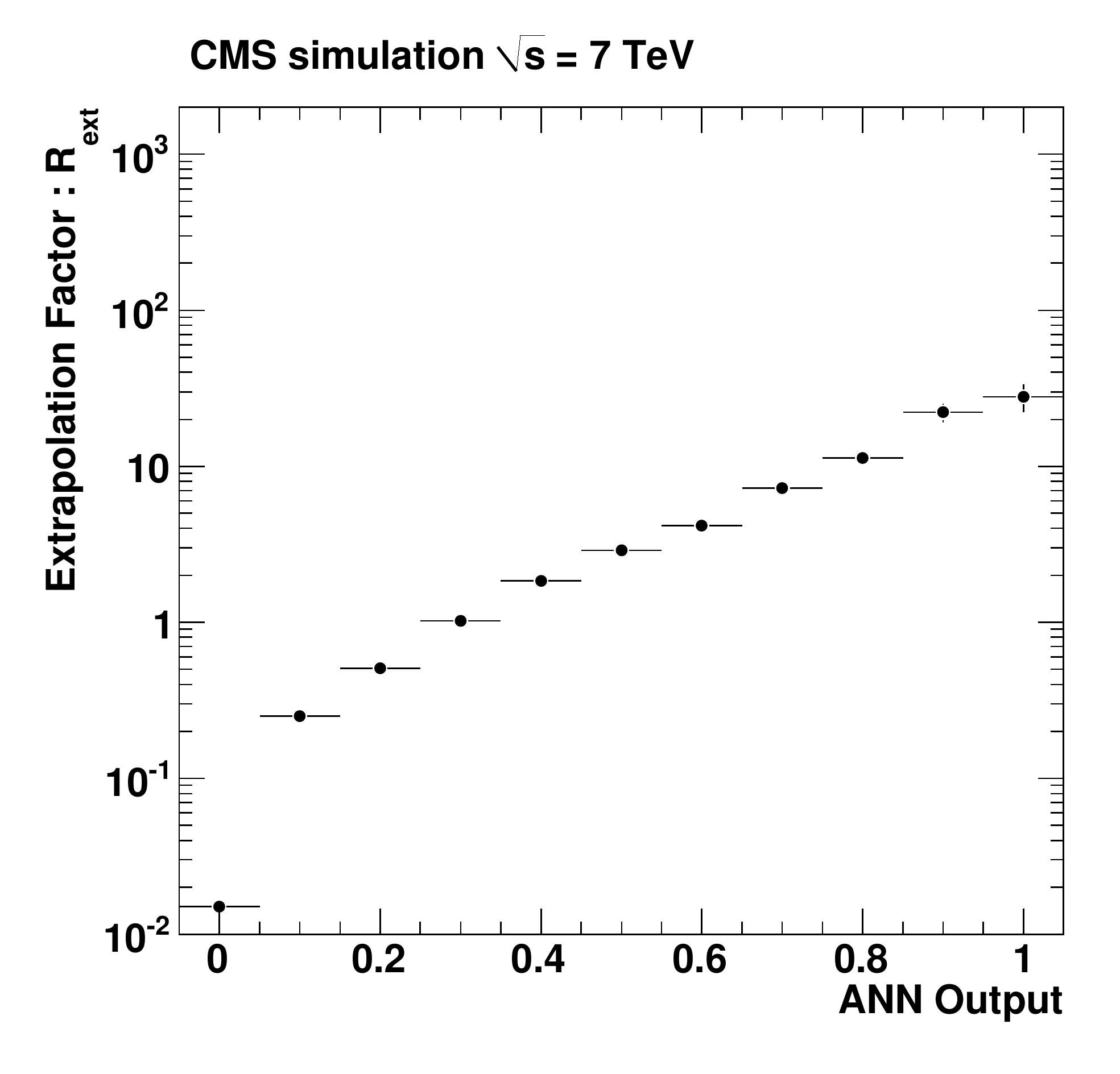}
\caption{ Extrapolation factor $R_\text{ext}$, as obtained from simulated
events.}
\label{figExtra}
\end{figure}

Next, an extrapolation ratio,
$R_\text{Ext.}=\frac{\mathrm{ANN(SM)}^\text{MC}_\text{SR}}{\mathrm{ANN(SM)}^\text{MC}_\text{CR}}$
obtained from simulated events, is defined for each bin in
the ANN output distribution as the ANN output for the
SM-only hypothesis in the signal region divided by the ANN output
for the SM-only hypothesis in the control region. The extrapolation
factor, $R_\text{ext}$, exhibits a smooth monotonic behavior, as shown in Fig. \ref{figExtra}.

Finally, the ANN output from data in the control region, where
only SM physics is assumed to be present, is multiplied by
the extrapolation factor, $R_\text{ext}$, to predict
the ANN output SM in the signal region, $\mathrm{ANN
(SM)}^\text{prediction}_\text{SR}$:
\begin{equation}
 \mathrm{ANN(SM)}^\text{prediction}_\text{SR}=\mathrm{ANN(SM)}^\text{data}_\text{CR} \times
\frac{\mathrm{ANN(SM)}^\text{MC}_\text{SR}}{\mathrm{ANN(SM)}^\text{MC}_\text{CR}}.
\end{equation}
The primary control region is further subdivided into a \ttbar enriched
one with $\ETslash >30\GeV$ and
$M_{\ell\ell}  \notin [75,105]$\GeV, denoted as ``control region A'', and
separately into a $\cPZ+\text{jets}$
enriched one with  $\ETslash <30\GeV$ or  $75\GeV  < M_{\ell\ell} <105\GeV$,
denoted as ``control region B''.
These are not used in   the analysis. However they  provide quality control cross-checks
(level of agreement between  data and simulation) for the two main backgrounds
that affect the analysis.

\begin{figure*}[hbtp]
\begin{center}
\includegraphics[width=0.48\textwidth]{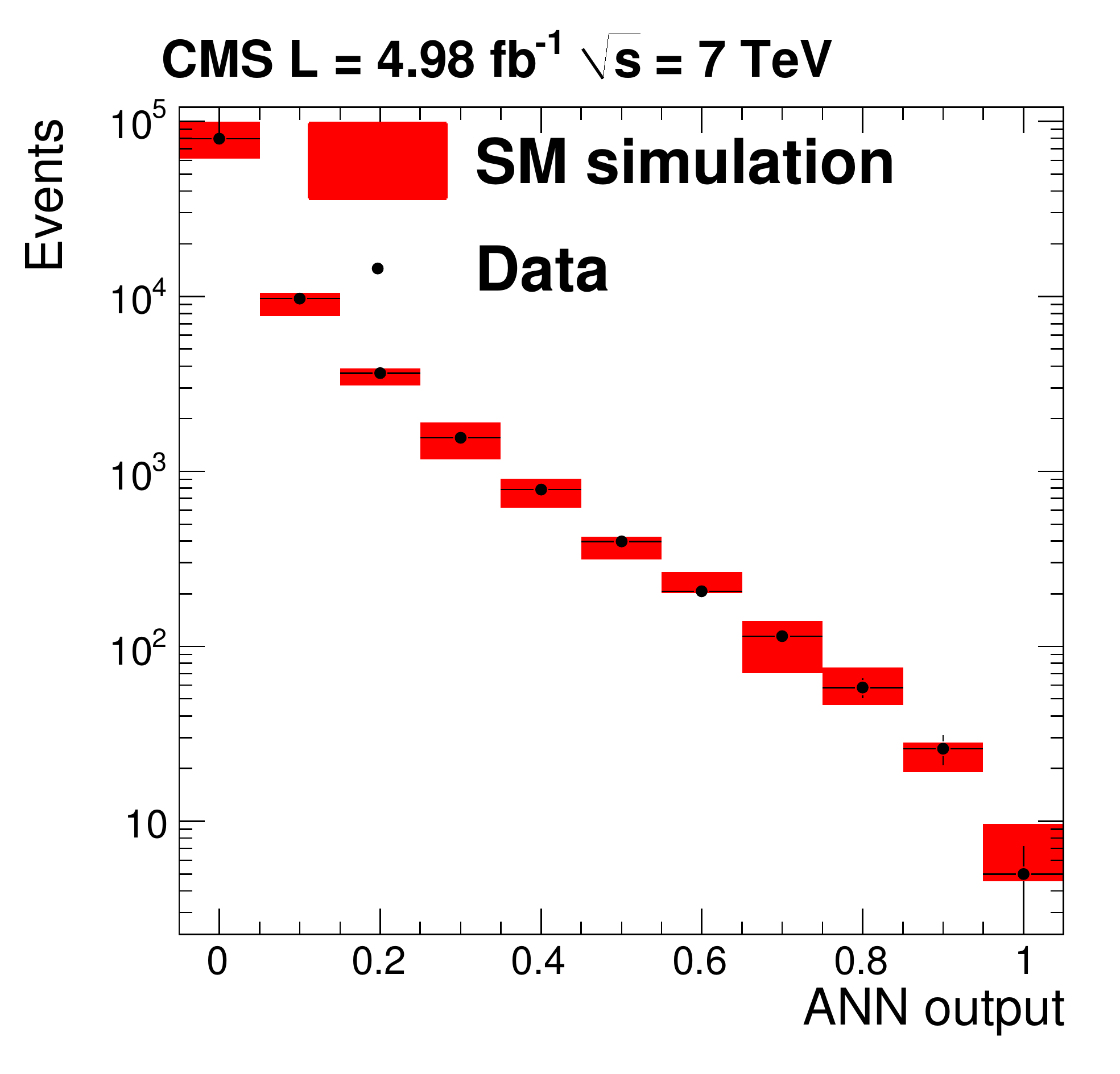}
\includegraphics[width=0.48\textwidth]{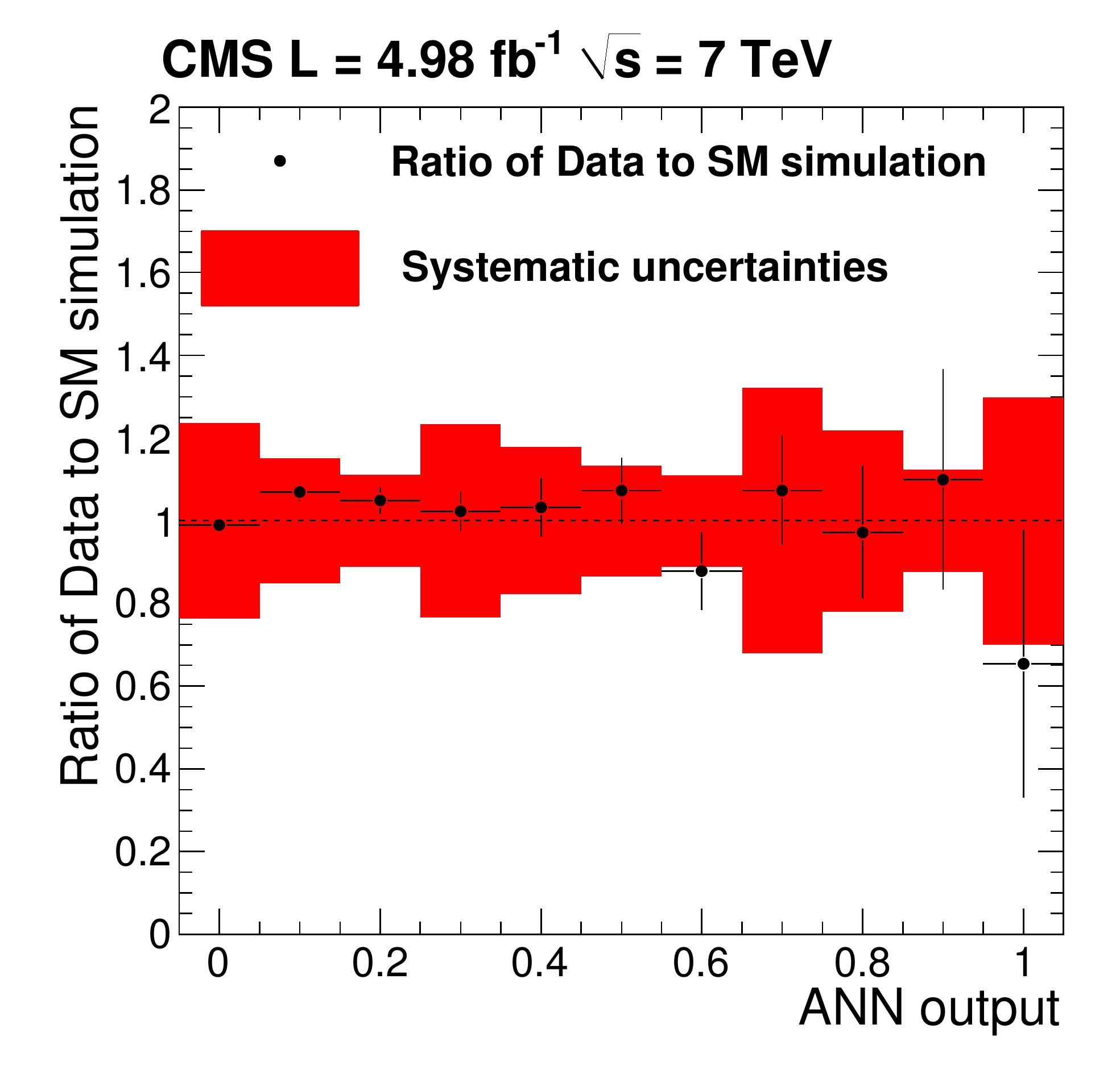}
\includegraphics[width=0.48\textwidth]{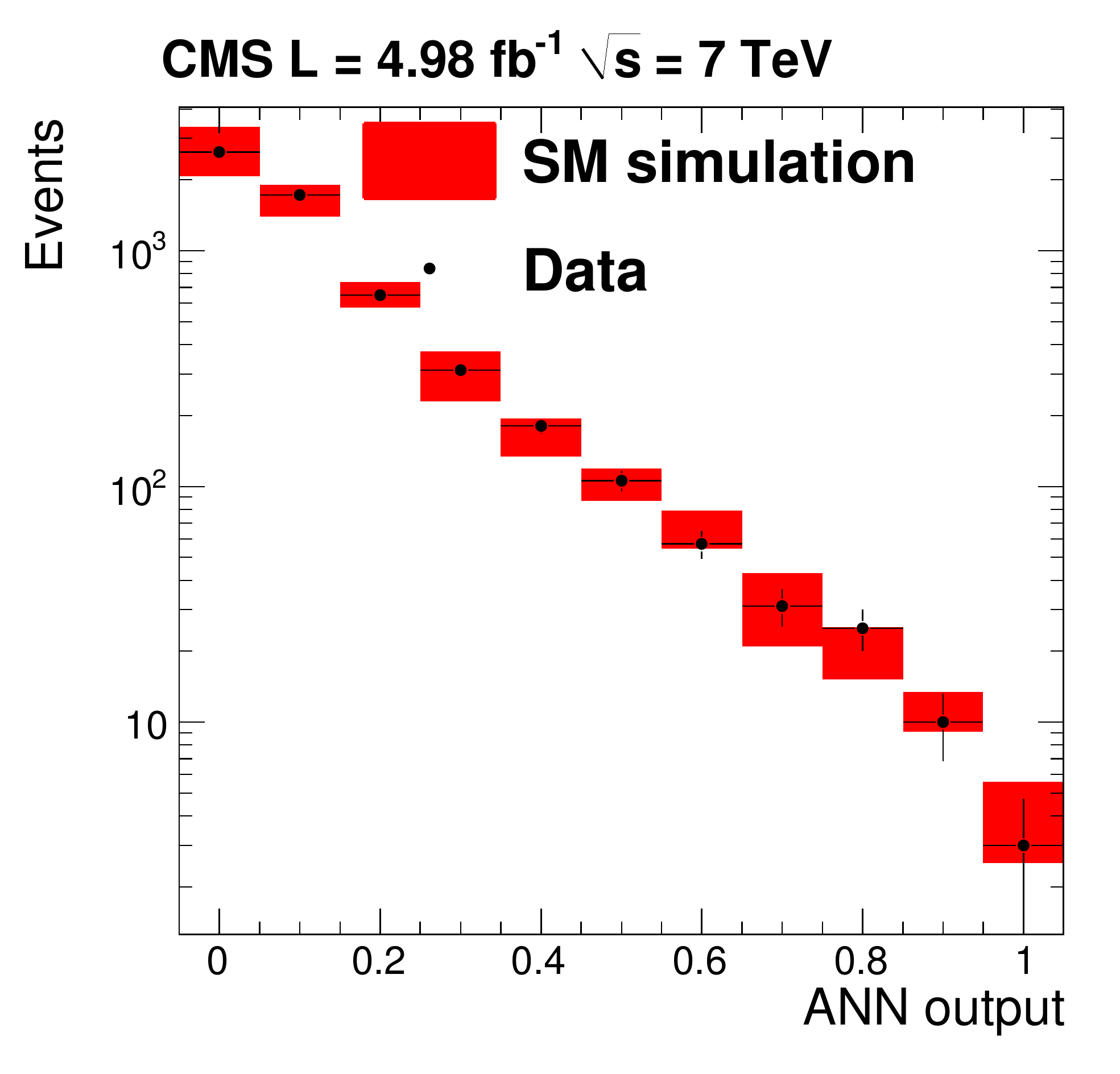}
\includegraphics[width=0.48\textwidth]{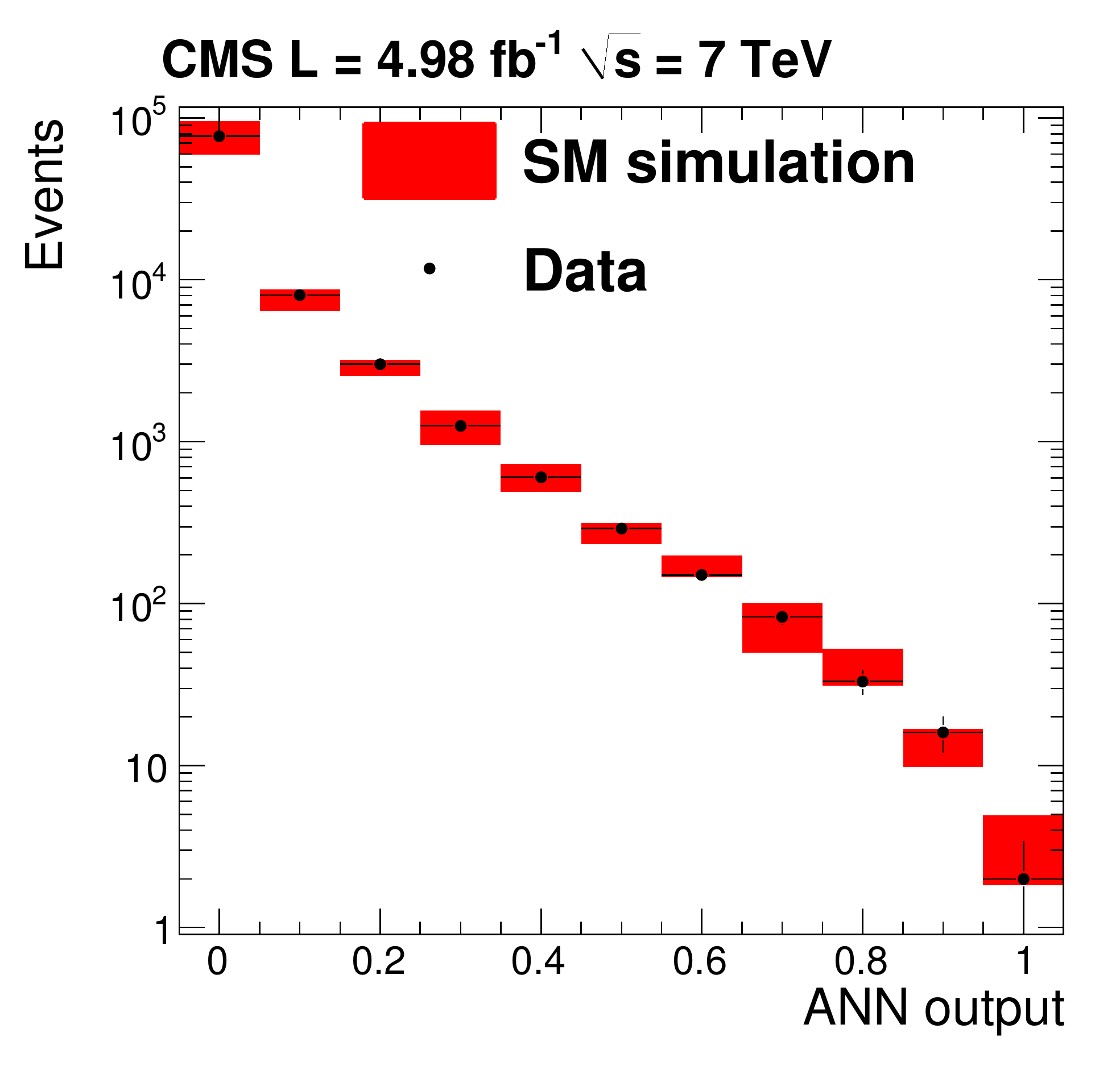}
\caption{Data (black points with error bars) vs. simulated events (red bands) comparison of the
 ANN output
 distributions in the various control regions. Top: The ANN output in the control region
used to perform the extrapolation  with systematic uncertainties included
(left), and the ratio  between data and simulated events (right) with both statistical
(black error bars) and systematic (red bands)  uncertainties shown.
Bottom: The main control region is subdivided into the following two to perform cross-checks:
\ttbar enriched
control region A (left), $\cPZ+\text{jet}$ enriched control region B (right) with systematic
uncertainties included.}
\label{ref:fig20}
\end{center}
\end{figure*}

Figure \ref{ref:fig20} compares the ANN output distributions of  data and
simulated events  in the control regions as defined above.
Agreement between data and simulation is observed both in the primary
control region used to define the ANN output, as well as in the \ttbar
and $\cPZ+\text{jets}$ dominated control regions ``A'' and ``B''.

Similar agreement between  data and simulation for the ANN input variables in the control region is
observed as well. This helps to  confirm that the simulation is appropriate
to train the ANN and adequate to be used for the estimation
of systematic uncertainties.
\section{Systematic Uncertainties}

Systematic uncertainties of the ANN output prediction for the
SM-only hypothesis, obtained as described in Section~\ref{SMTemplate},
are estimated with simulated data using the following
procedure. A systematic effect is introduced into the
simulated data for all events in the sample before any
preselection is applied. The nominal SM extrapolation factor $R_\text{ext}$ is then used to obtain a
new ANN output prediction for the signal region  corresponding to the
systematic effect under study. Next, the ANN output prediction, corresponding to the
systematic alteration, is compared against the ANN output for the original
sample, without any systematic effects introduced. A binned ANN output distribution is studied for this
analysis. The
relative difference in ANN outputs for each bin, is assigned as a
bin-by-bin systematic uncertainty. Similarly, the relative difference in the integrated number of events
above a certain ANN output is assigned as a systematic uncertainty to the number of signal-like events.
Finally, for each bin, the relative differences for all
systematic effects studied are added in quadrature. This
results in a bin-by-bin total systematic uncertainty in
the ANN output prediction. In a similar manner the relative differences in the integrated
number of events above some ANN value are added in quadrature  yielding the total systematic
uncertainty on the number of signal-like events.

 The overall systematic uncertainties corresponding to the seven input
 variables used for the ANN construction, as well  as the uncertainties in the cross sections of
 the SM backgrounds, are shown in Table \ref{table7} for the ANN optimal
 selection.

\begin{table*}[htbH]
\begin{center}
\footnotesize
\topcaption{Systematic uncertainties considered in the predicted background,
along with their magnitude, and the impact they have on the final ANN output
prediction when the signal selection requirement at 0.95 is applied.}
\begin{scotch}{ccc}
 Quantity & Syst. uncertainty & Syst. error (\%) on the SM prediction \\ \hline
 Missing transverse energy  \ETslash &     ${\pm} 10 \%$ & 26        \\
 Leading, subleading   jet $\pt$     &     ${\pm} 3 \%$  & 10        \\
 $\frac{\sum{\ET^\text{lepton}}}{\sum{\ET}}$&   ${\pm} 2 \%$  &  9        \\
 Transverse mass $\mt$               &     ${\pm} 5\%$  &  6        \\
 Dilepton mass $M_{\ell\ell}$              &     ${\pm} 1 \%$  &  1        \\
 Number of jets $N_\text{jets}$           &     ${\pm} 0.5\%$ & $<$1      \\ \hline
 \ttbar cross section            &     ${\pm} 4\%$  &  $<$1        \\
 QCD, $\PW+\text{jets}$, $\cPZ+\text{jets}$ cross sections &   ${\pm}50\%$,${\pm} 3\%$,${\pm} 3\%$  & $<$1      \\
 Total                                 &                 & 30        \\
\end{scotch}
\label{table7}
\end{center}
\end{table*}

The magnitude of the systematic alterations for the jet energy scale is
taken from dedicated CMS measurements \cite{ref:jecp}.
While the clustered energy scale of \ETslash is known to the $3\%$ level
in CMS and the unclustered energy scale for  \ETslash is known to within
$10\%$~\cite{ref:metp}, this analysis uses a conservative $10\%$ for the
overall \ETslash  systematic uncertainty.

For the input ANN variables for which there is no dedicated CMS measurement, the level of agreement
between data and simulation in the control region is used to obtain an estimate of the systematic uncertainty.
Therefore, the control region is used to constrain the systematic uncertainties in these cases.
Given the above, the difference between data and simulation for the migration of events from
the one-jet to the two-jet bin is  estimated to be 0.5\%. Similarly, the systematic uncertainty on the ratio
of the lepton  to the total transverse energy is estimated to be 2\%, and the $\mt$ uncertainty is
estimated to be 5\%.
The dilepton mass scale uncertainty of 1\% is taken from the CMS measurements of the $\cPZ$ peak \cite{ref:ZJET}.

The relative fraction of \ttbar and $\cPZ+\text{jets}$ backgrounds is
observed to vary as a function of the ANN output, as well as across the signal and control regions.
In order to account for any remaining differences,
the cross sections of all background components are left to vary within their
uncertainties, taken from the recent CMS measurements for the \ttbar
\cite{ref:ttbar} cross section, and using a conservative 50\% uncertainty on the  QCD cross section.
The $\cPZ+\text{jet}$ cross section uncertainty ($<$3\%)  \cite{ref:ZJET}, and the $\PW+\text{jet}$ cross section uncertainty $<$3\%)~\cite{ref:ZJET}  produce a negligible systematic effect on the ANN output.

The systematic uncertainties associated with the signal acceptance and
 efficiency (ANN selection), along with  their magnitude, are summarized in Table  \ref{ref:table12}.
\begin{table}[htbH]
\begin{center}
\footnotesize
\topcaption{Systematic uncertainties on signal acceptance and efficiency.}
\begin{scotch}{cc}
Source of systematic & Uncertainty  \\  \hline
Lepton triggers ($\pt>20\GeV{}$) & 3\%  \\
Lepton isolation & 5\% \\
Integrated luminosity & 2.2\% \\
ANN selection & 17\% \\
Total & 18\% \\
\end{scotch}
\label{ref:table12}
\end{center}
\end{table}
 The uncertainty on the lepton triggers and the lepton isolation are the
 same as the ones estimated in Ref. ~\cite{ref:ref_013a}.  The relative ANN uncertainty
 for the signal is lower than the corresponding uncertainty for the background, due mainly to
 the different  ANN shapes for   these two populations (signal and background).
\section{Performance of the ANN}

\begin{figure}[h]
\centering
\includegraphics[width=0.48\textwidth]{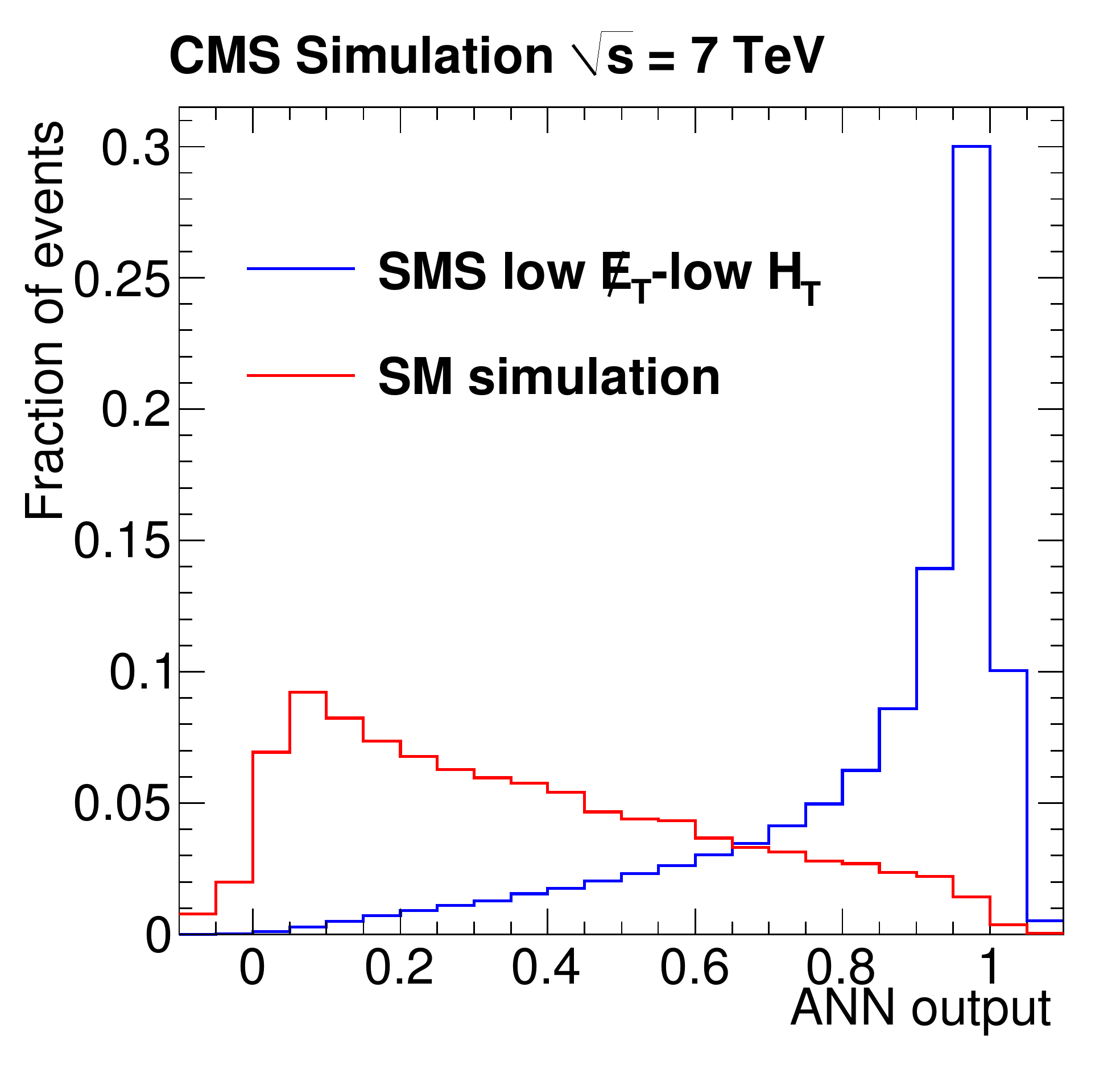}
\includegraphics[width=0.48\textwidth]{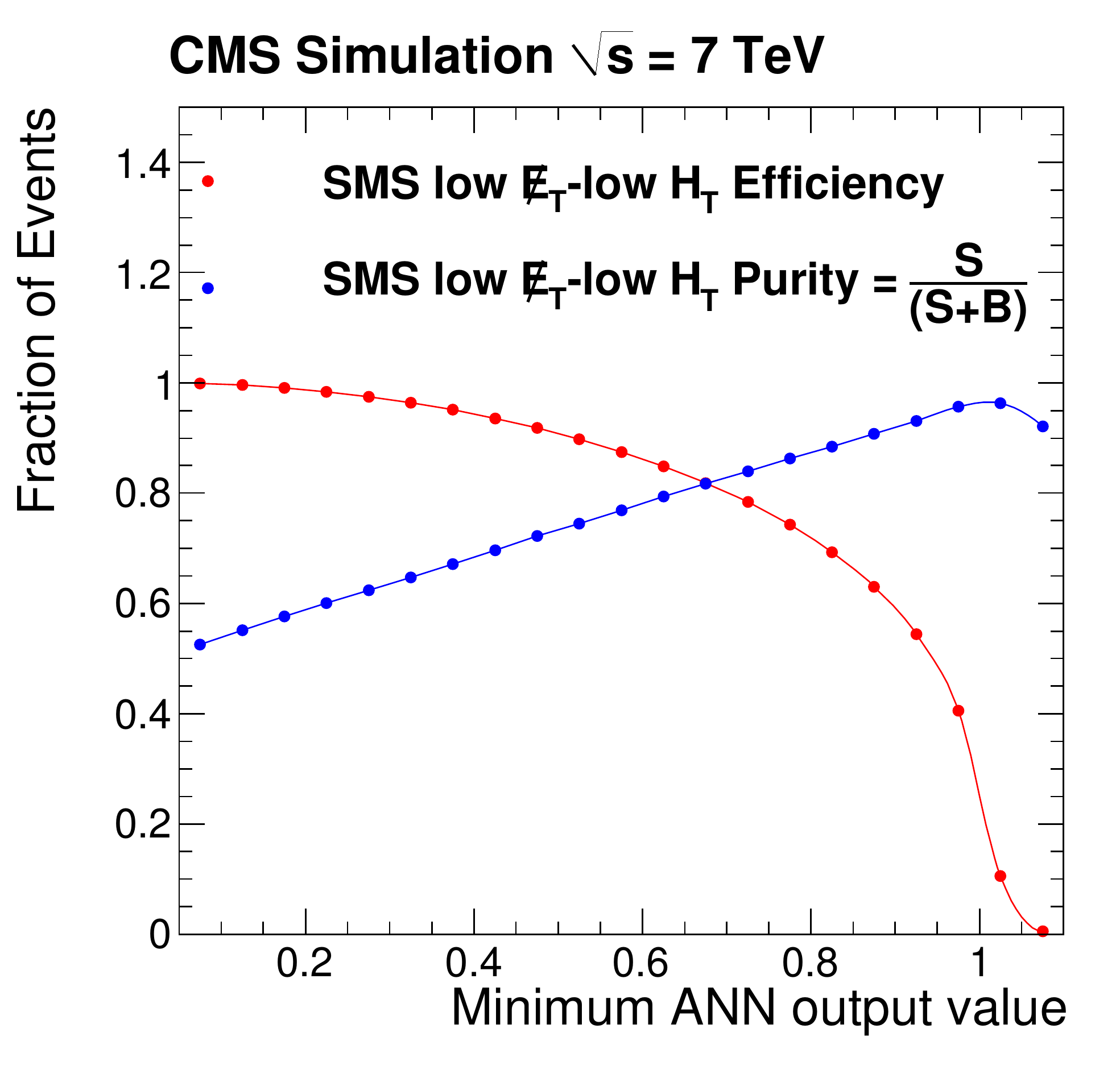}
\caption{ \cmsLeft: The  ANN output for the SM background (red bands) and SMS low \ETslash-low \HT
events  (blue line).  \cmsRight: Efficiency (red) and purity (blue) vs. the minimum ANN output value for SMS
low \ETslash-low \HT events in the signal region.}
\label{ref:fig10}
\end{figure}

The ANN output after the training is shown in Fig. \ref{ref:fig10} for
the signal (blue) and SM background (red) samples; the efficiency and
purity of the selected samples are also shown as a function of the ANN
output requirement.

When statistical and systematic uncertainties are taken into
account, the ANN output requirement yielding the best expected exclusion
limit in the SMS plane is $\mathrm{ANN}>0.95$. The expected number of SM and
signal events for the CMSSM
benchmark point LM6 after imposing the ANN  output requirement of $>$0.95 are shown in Table
\ref{ref:table4}. The remaining backgrounds are dominated by \ttbar events
in the dilepton final state, followed by $\cPZ+\text{jets}$ production at a much smaller level.

\begin{table}[htbH]
\begin{center}
\footnotesize
\caption{Expected number of events for signal and SM
background and for the ANN output greater than $>0.95$.
The NLO cross section is used for the CMSSM benchmark point LM6.}
\begin{scotch}{cc}
 Sample &  Events with ANN $>0.95$ \\ \hline
 \ttbar    & 125    $\pm 38$      \\
 $\cPZ+\text{jets}$      &  14   $\pm$  4      \\
 $\PW+\text{jets}$, \PW\PW, \PW\cPZ, \cPZ\cPZ, QCD & $<$1 \\
 Total SM Bkg. &  140  $\pm$ 42      \\
 LM6           &   40  $\pm$  1     \\
\end{scotch}
\label{ref:table4}
\end{center}
\end{table}

\section{Results}

\begin{figure*}[htbp]
\centering
\includegraphics[width=0.32\textwidth]{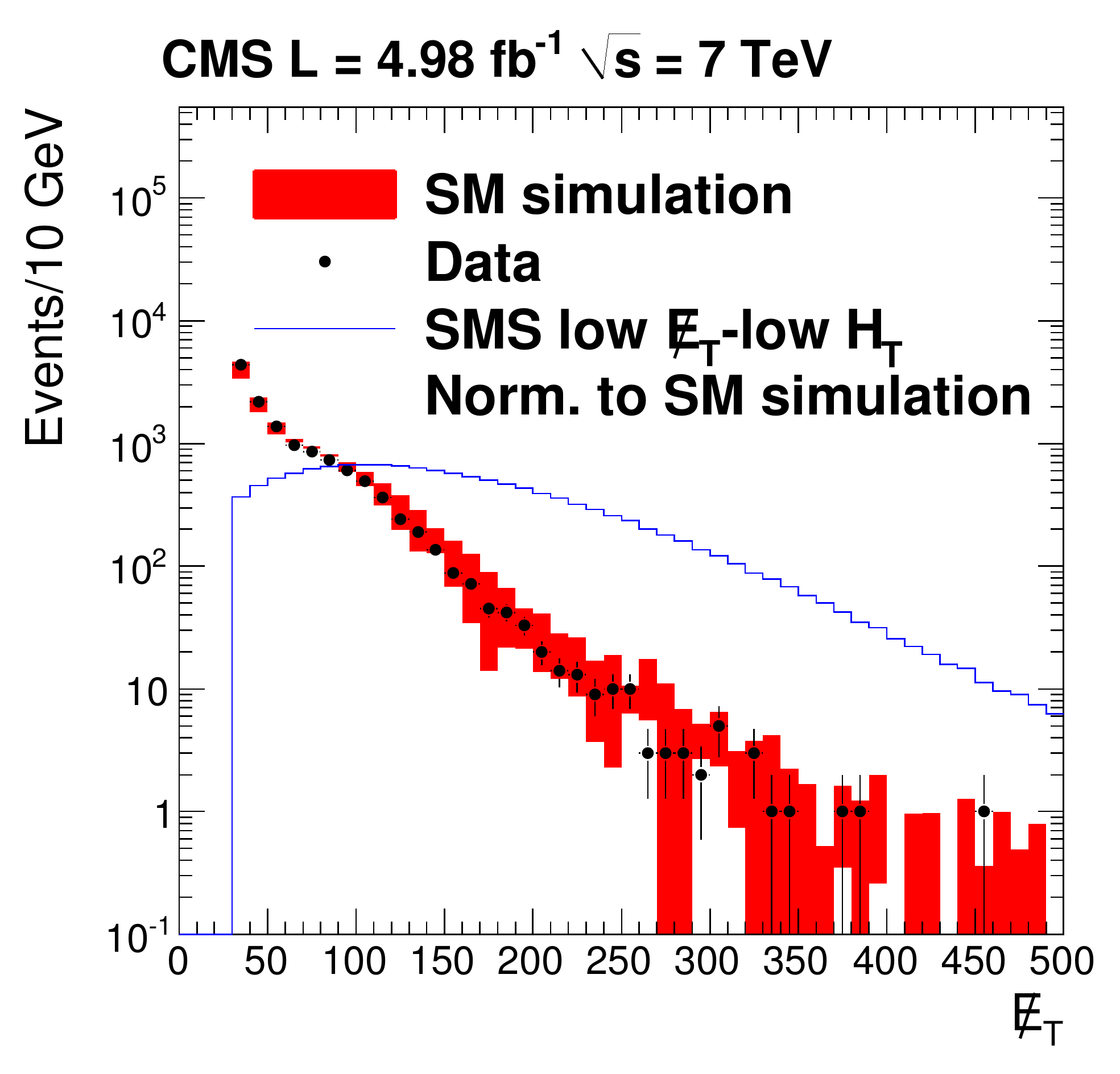}
\includegraphics[width=0.32\textwidth]{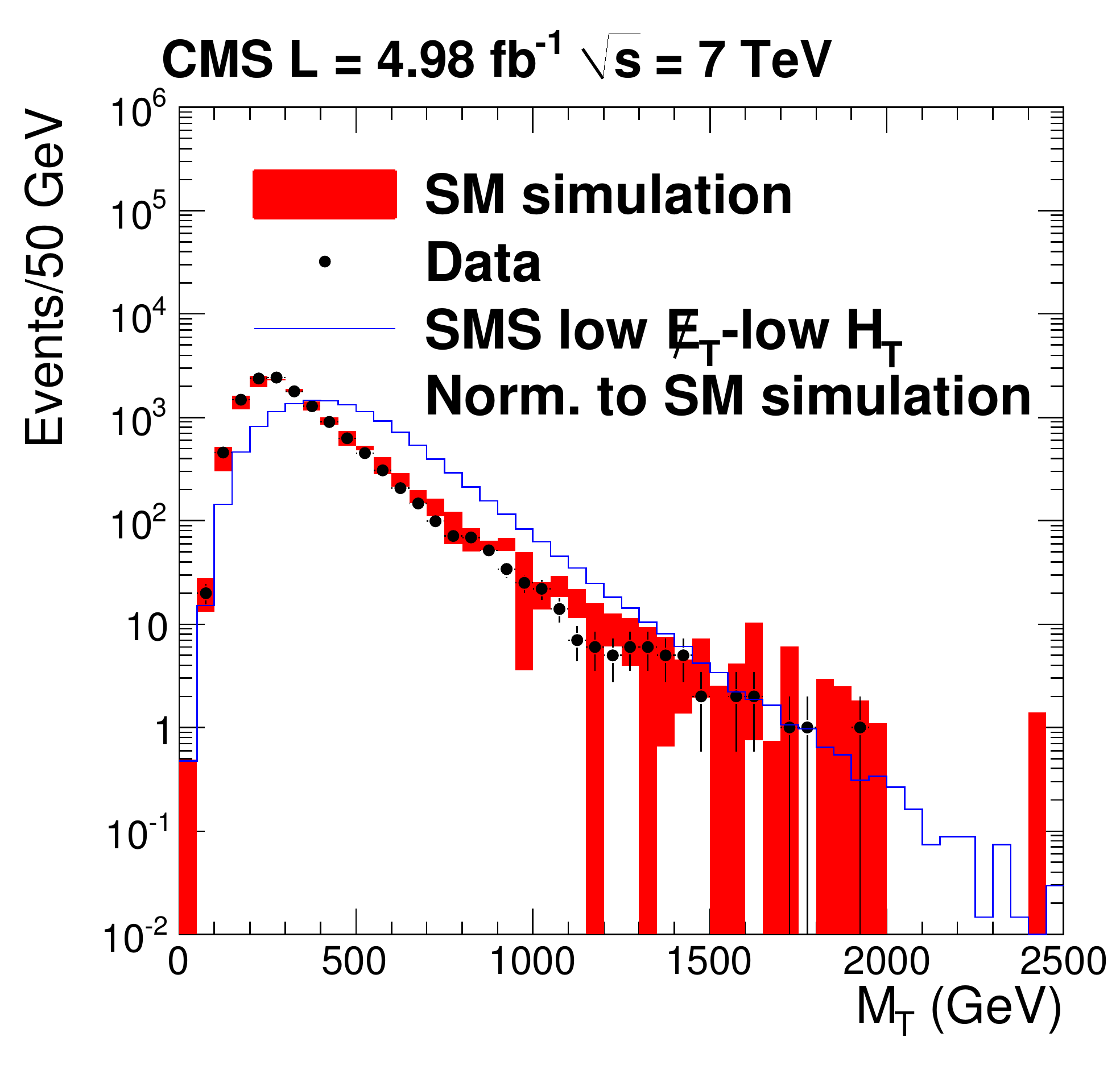}
\includegraphics[width=0.32\textwidth]{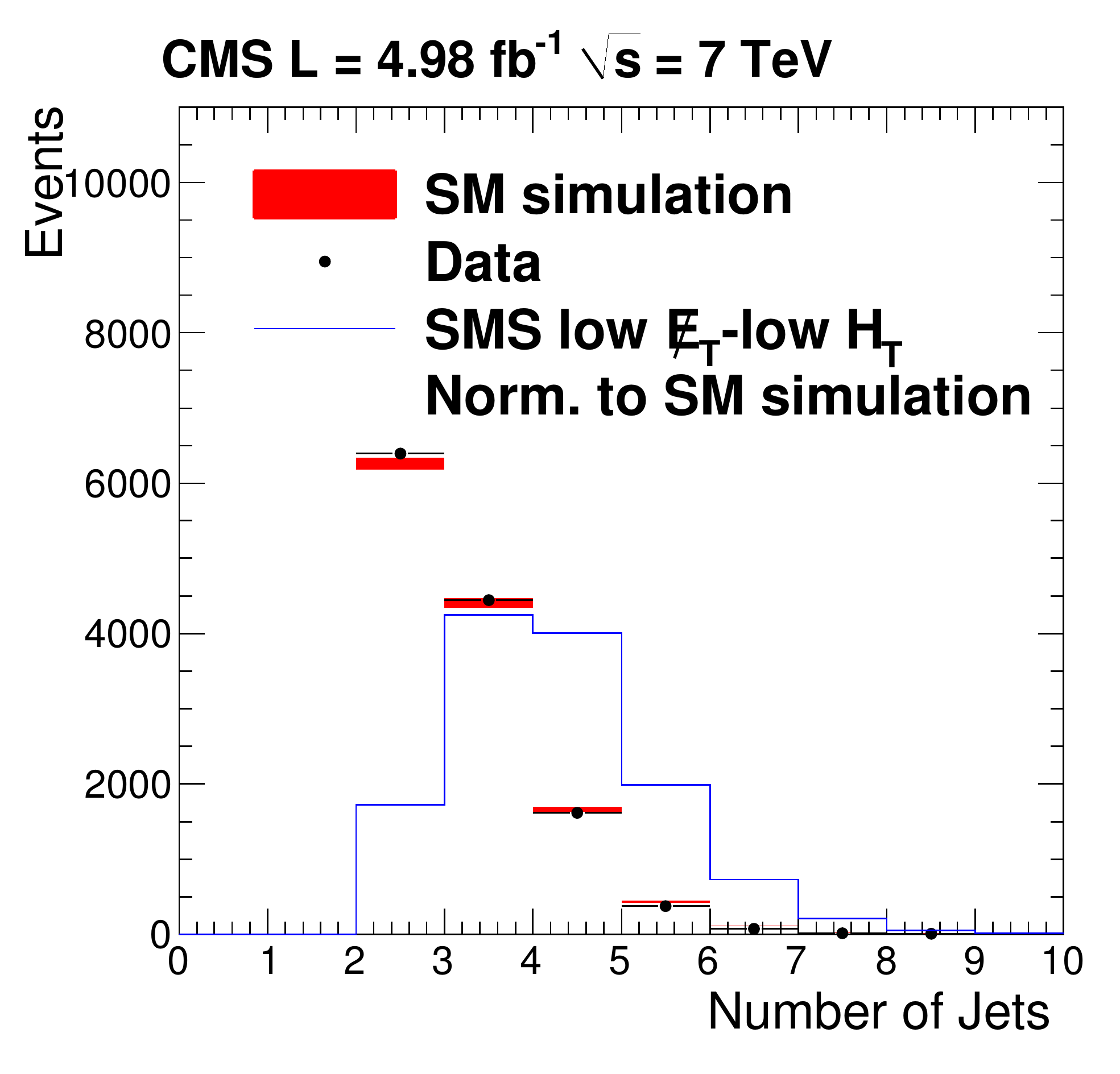}
\includegraphics[width=0.32\textwidth]{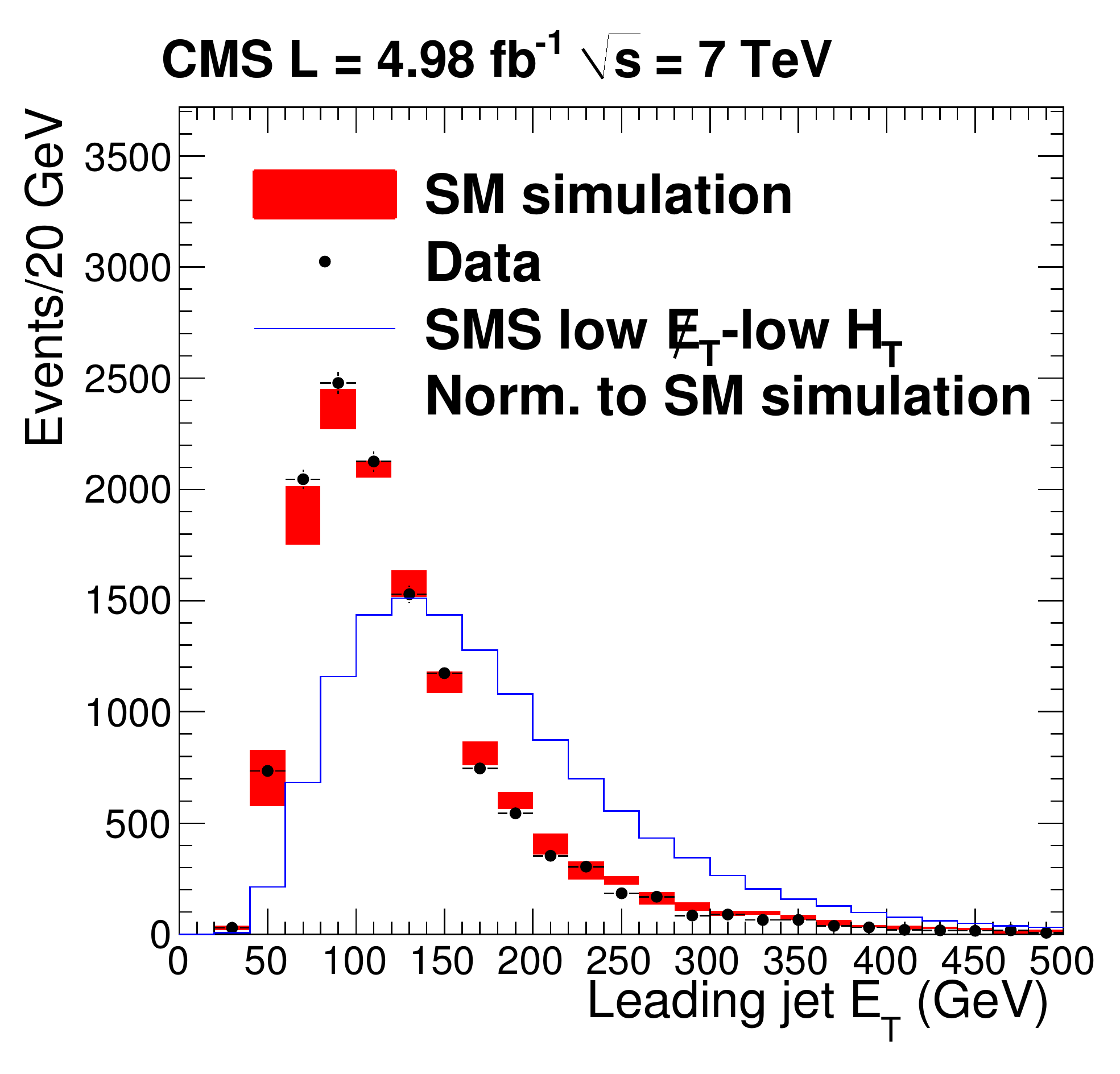}
\includegraphics[width=0.32\textwidth]{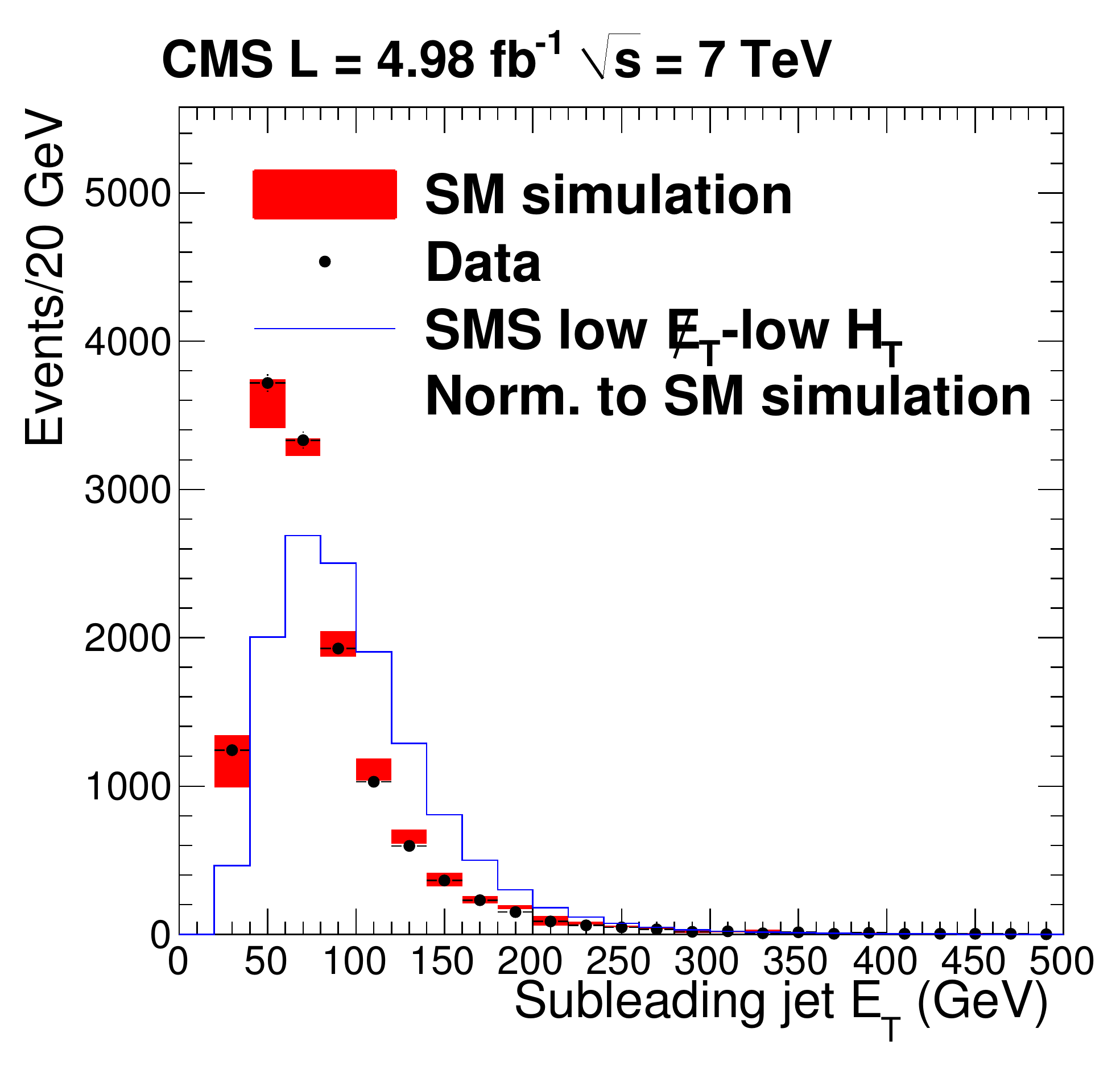}
\includegraphics[width=0.32\textwidth]{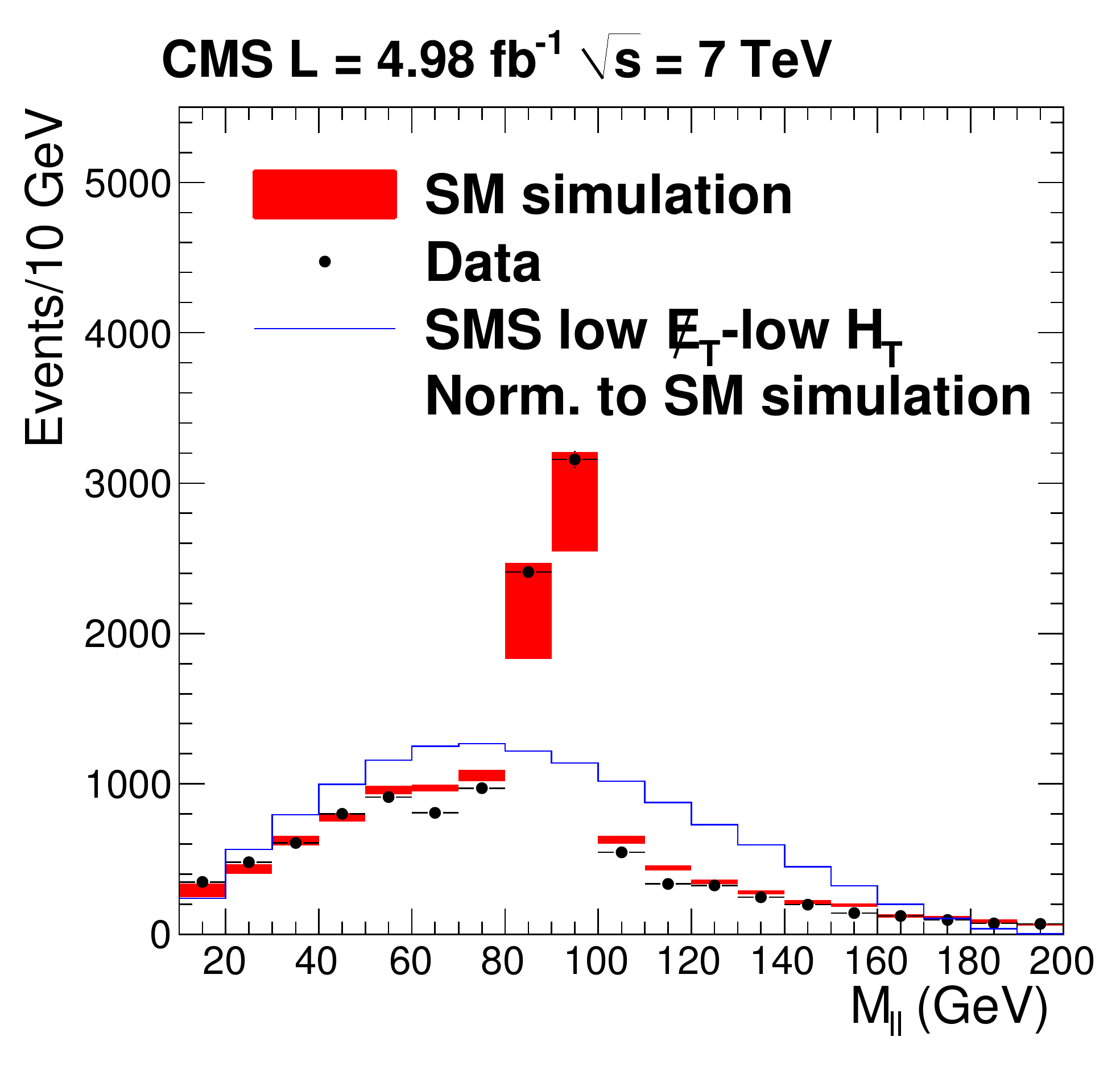}
\includegraphics[width=0.32\textwidth]{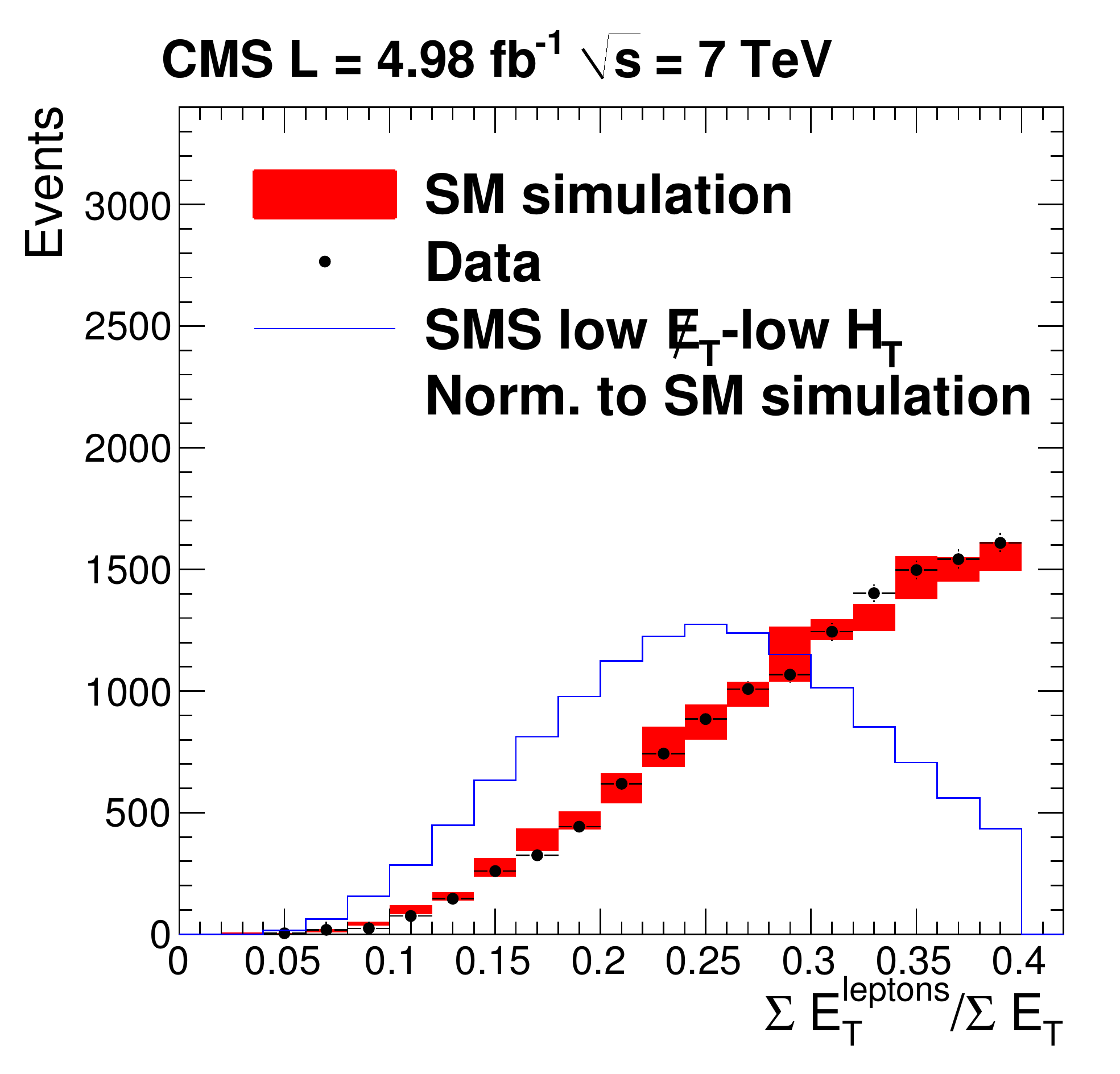}
\caption{ The distributions in seven input ANN variables for
 simulated SM background events (red shaded band showing combined statistical and systematic
 uncertainties) and SMS low-\ETslash low-\HT events (blue histogram), normalized to the same number of events. The data
 are also shown  for comparison (black points with error bars).}
\label{ref:fig9}
\end{figure*}

The seven input ANN variables are shown in Fig. \ref{ref:fig9} for simulated and data events,
after the candidate event selection criteria are applied and for signal events.
Data and simulation are consistent with each other, within the statistical and systematic
uncertainties.

\begin{figure}[hbtp]
   \begin{center}
\includegraphics[width=0.48\textwidth]{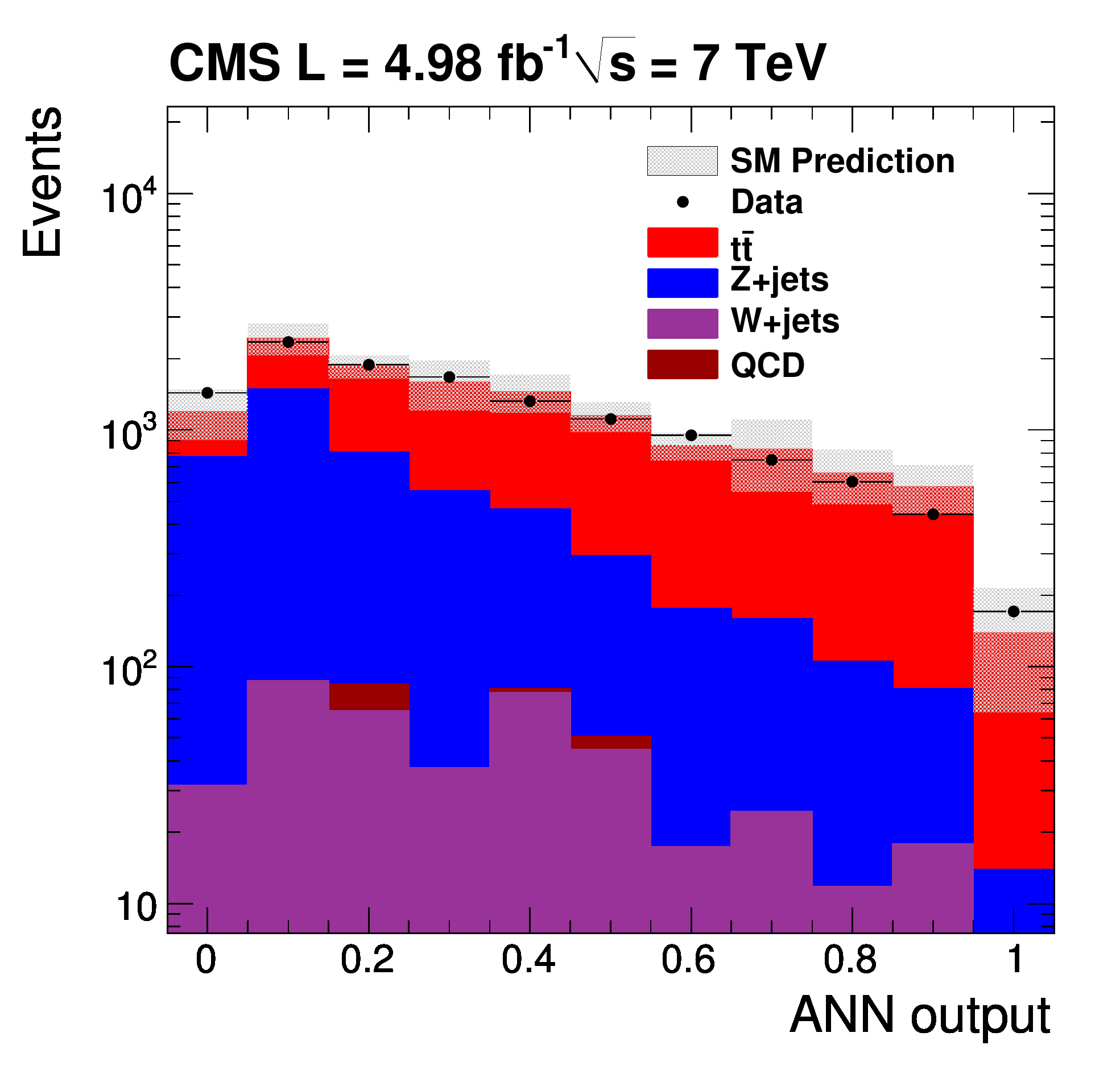}
     \caption{ The  ANN output for the data (black points) and  the SM prediction based on control regions in data
               (gray band) in the signal region. The uncertainty on the SM prediction includes both statistical
                and systematic components.}
      \label{ref:fig24}
   \end{center}
 \end{figure}

Figure \ref{ref:fig24} shows the comparison between the SM ANN prediction and the data in the signal
region including  statistical and systematic uncertainties.

\begin{figure*}[hbtp]
\begin{center}
\includegraphics[width=0.48\textwidth]{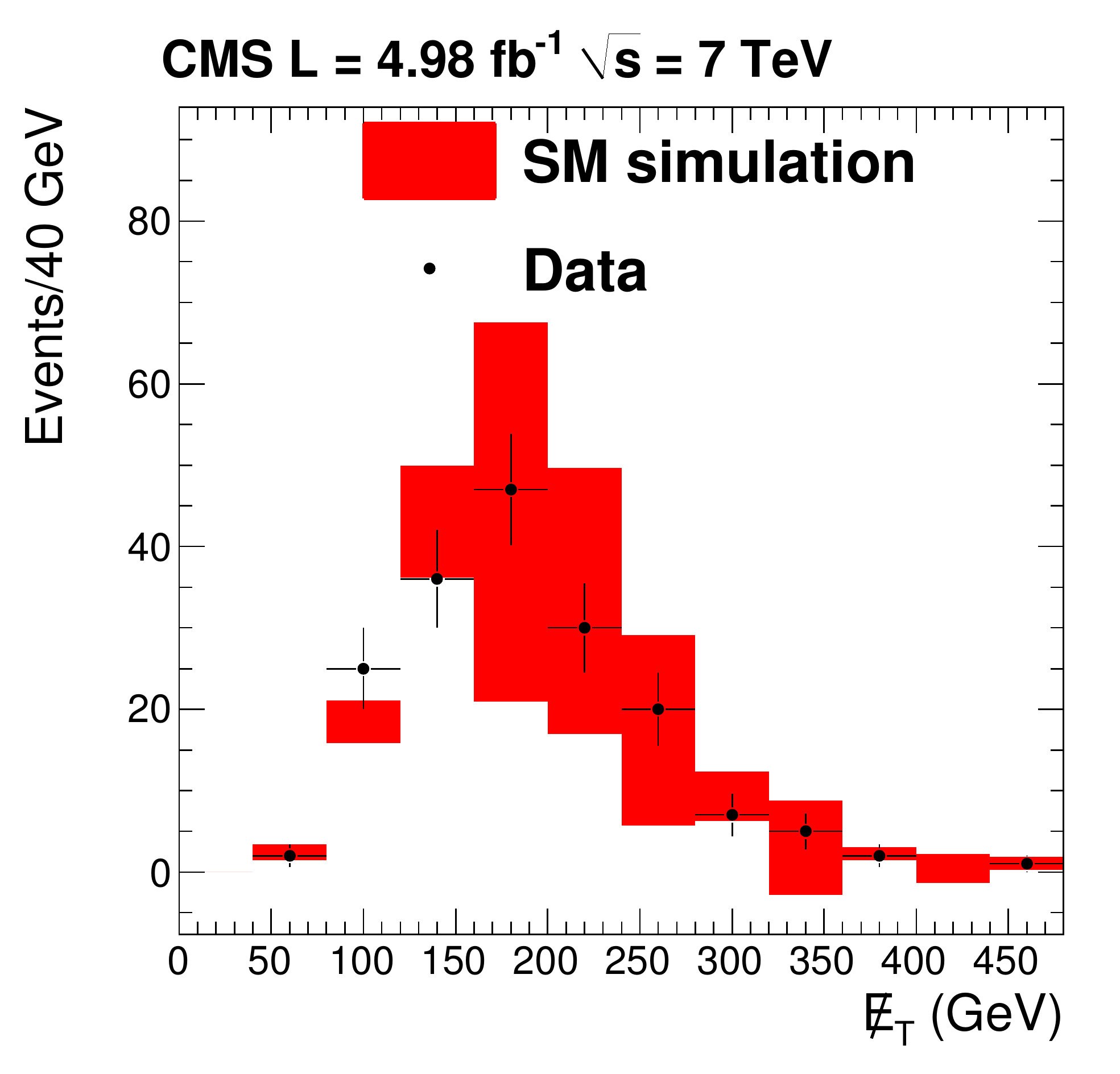}
\includegraphics[width=0.48\textwidth]{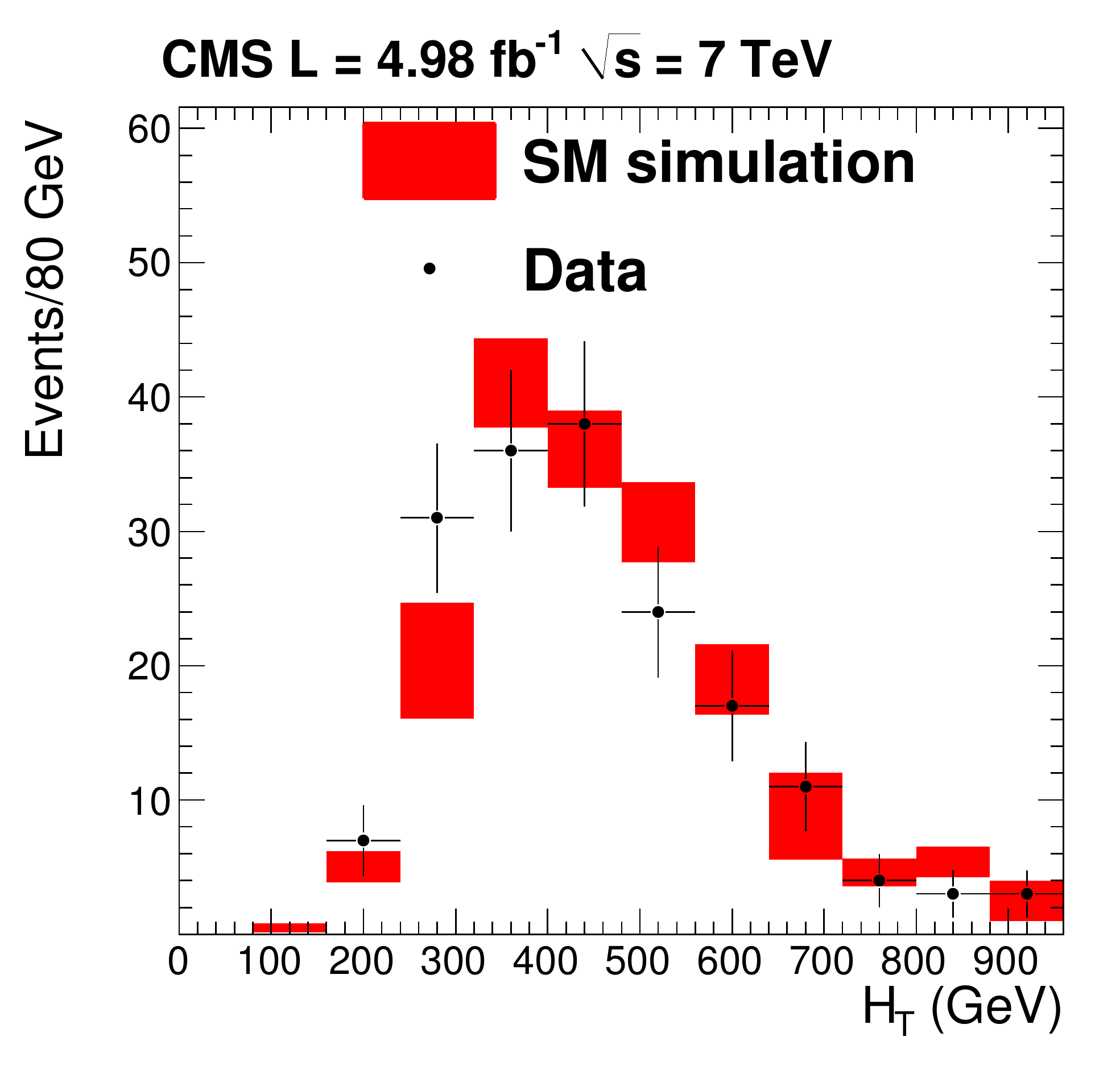}
\includegraphics[width=0.48\textwidth]{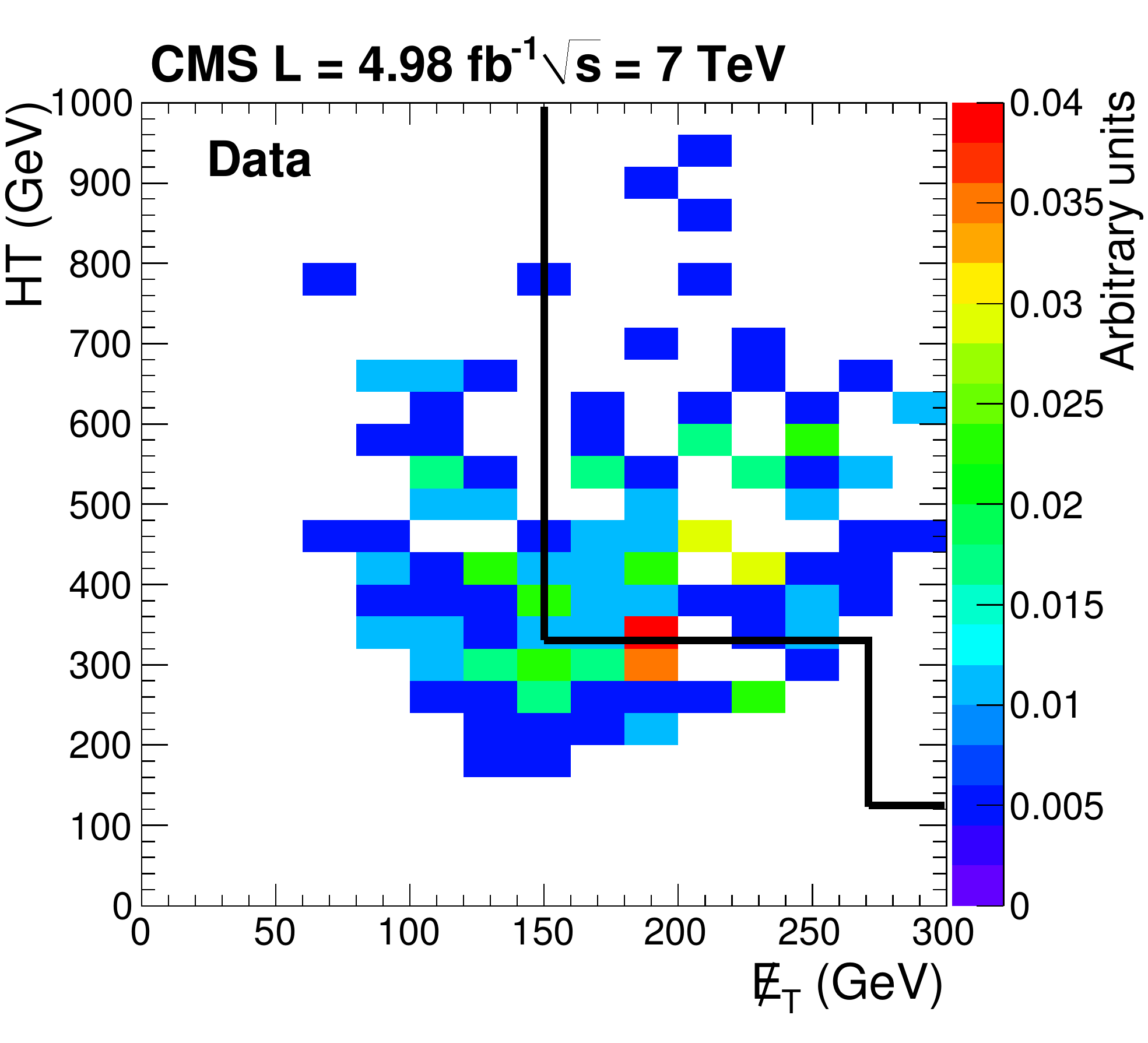}
\includegraphics[width=0.48\textwidth]{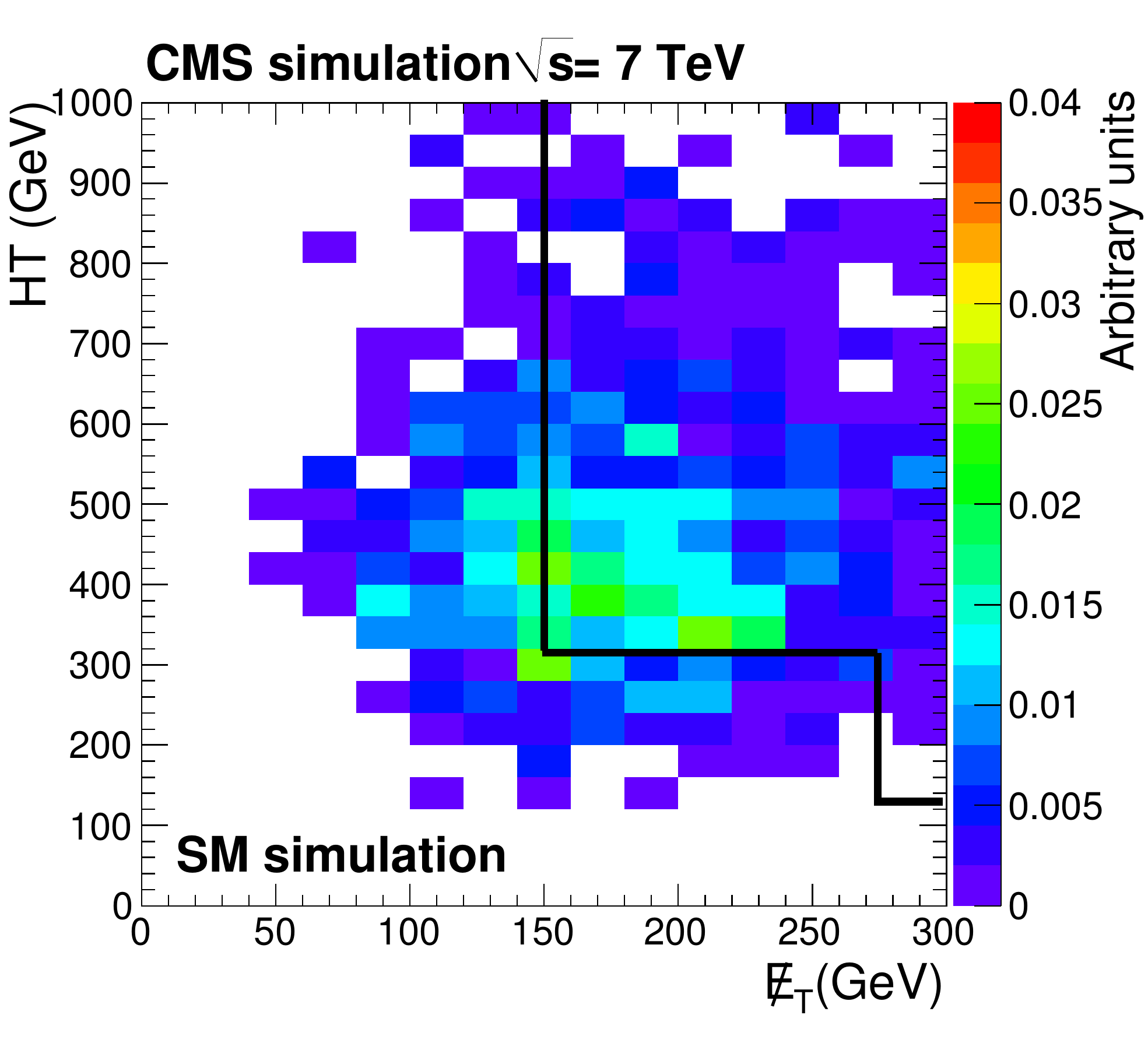}
\caption{Distribution in \ETslash (top left)  and \HT (top right) for signal-like events (ANN
output $>0.95$) for data (black points with error bars) and simulation (red shaded band).
The \ETslash vs. \HT distributions for signal-like events ($\mathrm{ANN}>0.95$) in data (bottom left)
and simulation (bottom right); the regions to the right and up of the black line are the
ones also explored by other CMS opposite-sign dilepton searches.}
\label{ref:fig20a}
\end{center}
\end{figure*}

 In the signal-like region there are 171 events observed and $140^{+73}_{-46}\stat\pm 42\syst$
expected. The statistical error on the expectation comes from
 the number of data events in the control region. The  95\% confidence level (CL) upper limit (UL) on
 number of
 signal events is estimated to be 95. There is  agreement between expectation and observation at a
 68\% CL. Figure \ref{ref:fig20a} shows the \ETslash and \HT
 distributions for data and simulated events in the signal-like region.
 These figures illustrate that this analysis accepts signal-like events with  \ETslash  as low as 40\GeV or \HT as low
 as 120\GeV{} --- regions not explored  yet by other CMS analyses.

Finally, the observed and expected number of events are translated into limits on SUSY parameter space.
The 95\% CL upper limits are computed using a hybrid  CL$_\text{s}$
method with profile likelihood
test statistics, and lognormal distributions for the background
expectation ~\cite{ref:STAT1,ref:STAT2}.
The uncertainties in the NLO+NLL cross sections from the parton distribution
functions \cite{ref:SUSY_CROSS, ref:SUSY_CROSS2,ref:SUSY_CROSS3,ref:SUSY_CROSS4,ref:SUSY_CROSS5}, the
choice of the factorization and
renormalization scale, and $\alpha_\text{S}$, are taken into account for each
point, and are  evaluated according to the PDF4LHC recommendation
\cite{ref:PDF4LHC}. A constant signal acceptance systematic uncertainty of 18\% is assumed for each
point.
As described previously, the contamination of the signal in the control region is negligible
and hence not taken into account in the limit setting.

\begin{figure}[hbtp]
\begin{center}
 \includegraphics[width=\cmsFigWidth]{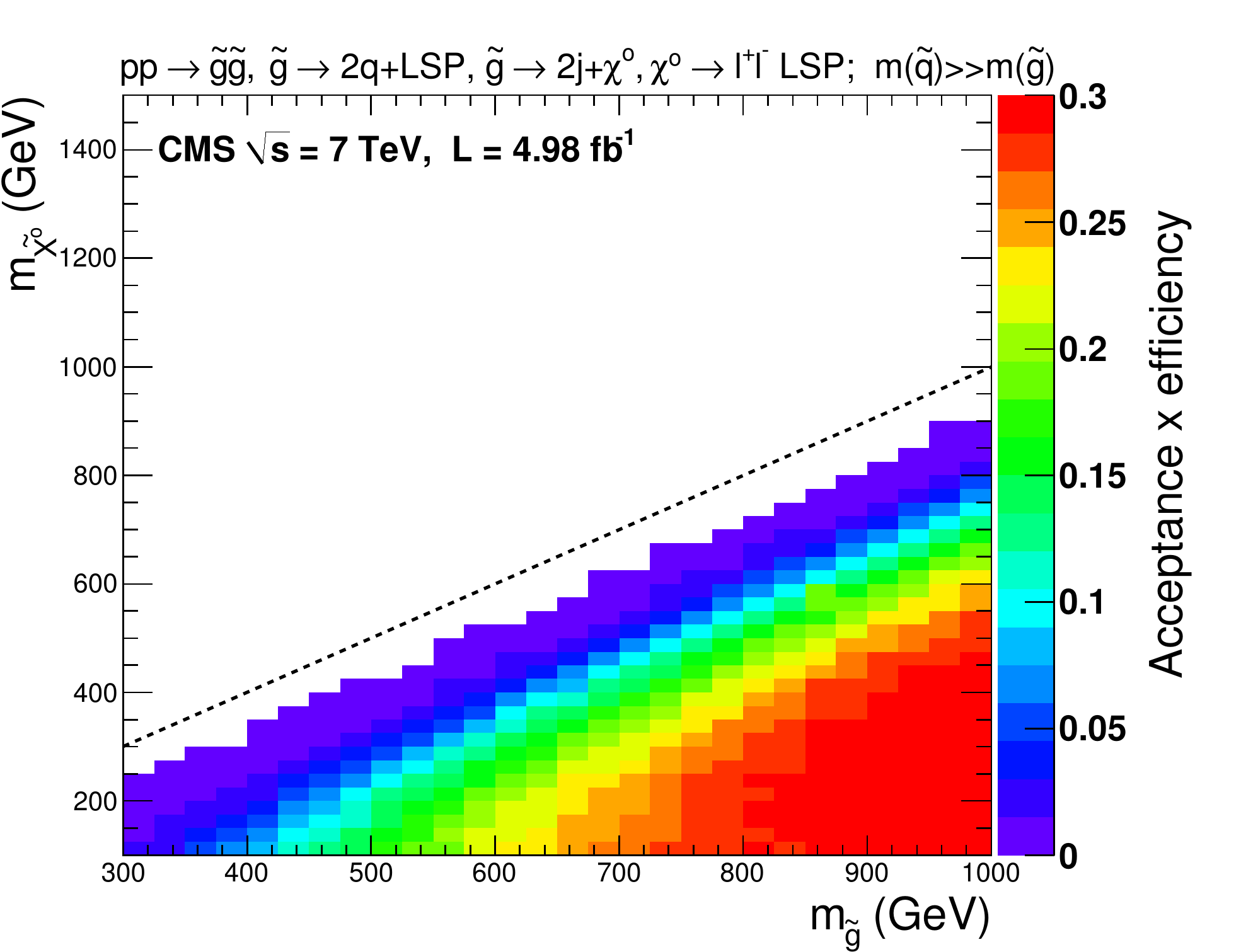}
 \includegraphics[width=\cmsFigWidth]{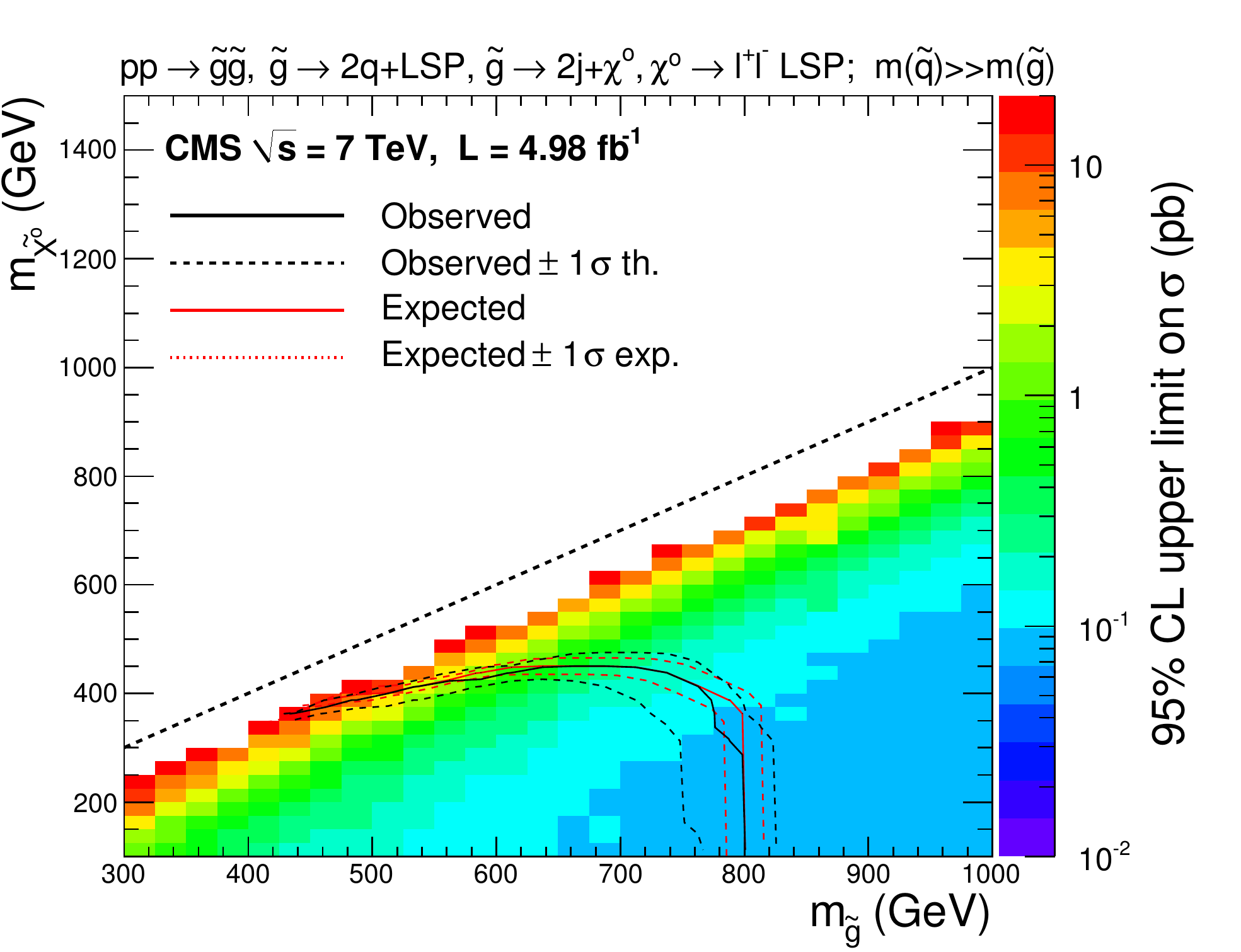}
 \caption{The 95$\%$ CL exclusion   limits on the simplified model scenarios with the ANN analysis.
  The acceptance (fraction of events surviving event selection and candidate event selection) $\times$ efficiency (fraction of events surviving ANN
  selection) (top) and $95\%$ CL upper
  cross
  section limit (bottom) are shown for different gluino and neutralino masses. The region just
  below the  diagonal is not considered due to inadequate initial state radiation modelling.}
      \label{figSensSM2n}
   \end{center}
 \end{figure}

\begin{figure*}[H!btp]
\begin{center}

\includegraphics[width=0.70\textwidth]{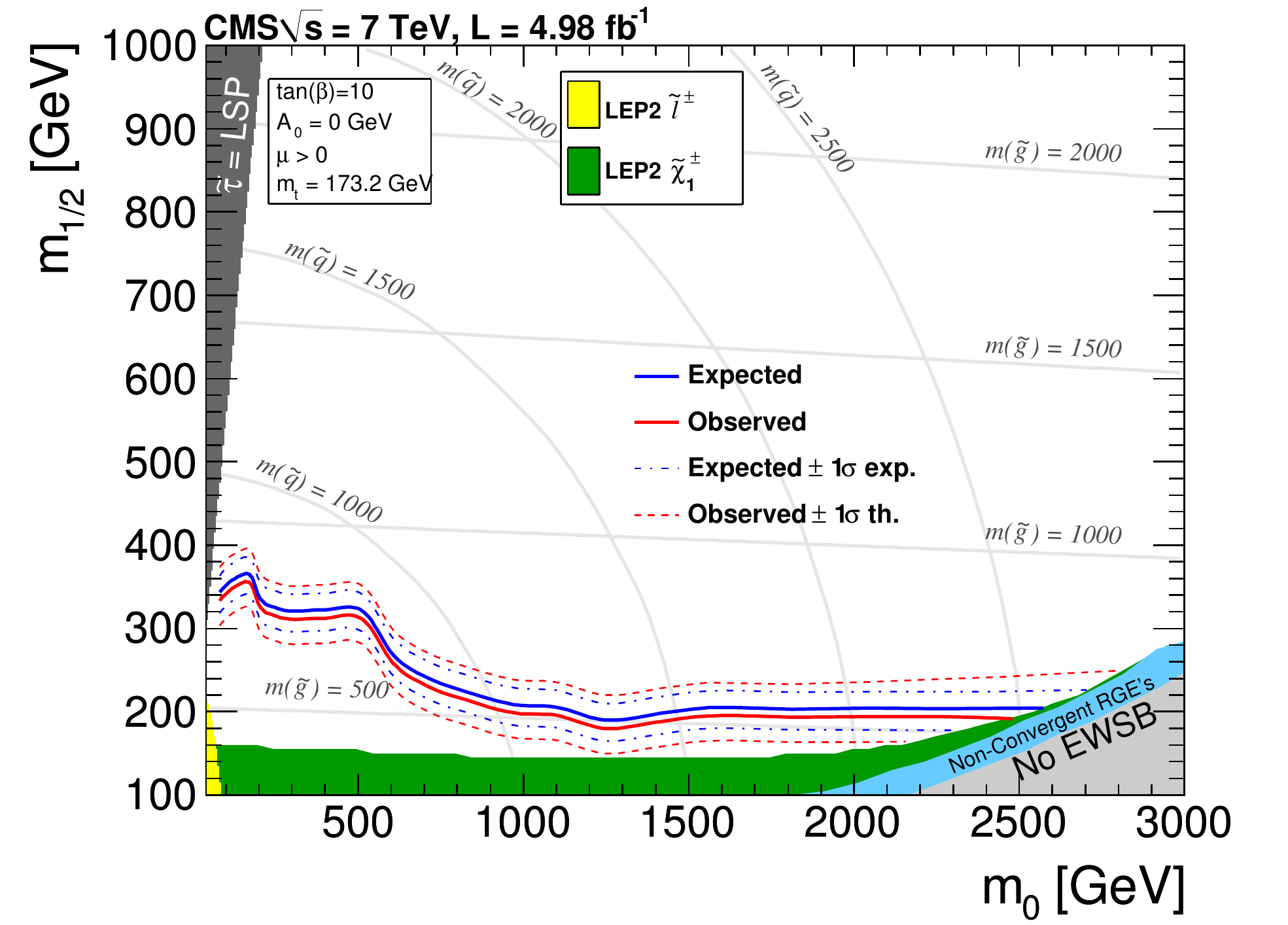}
\caption{ Expected (blue) and observed (red)  95\% CL exclusion
 limit for the ANN analysis (for ANN output $>0.95$) in the CMSSM plane. The one $\sigma$ experimental error
 around the expected limit, and the one $\sigma$ theoretical error around the observed limit are
  also shown.}
  \label{figSens}
  \end{center}
 \end{figure*}

The exclusion limits on SMS models are depicted in Fig. \ref{figSensSM2n}, and in the
 ($m_\text{0},m_\text{1/2}$) CMSSM
 plane are shown in Fig.  \ref{figSens} ~\cite{ref:ref_100}.

 As discussed earlier, for SUSY models that yield events with large
 \ETslash (CMSSM with $m_\text{0}<1000$ ), the ANN's performance is
 comparable to the data
 analyses using large \ETslash and \HT, and in some cases worse, given
 that the ANN has been trained with  models characterized by low
 \ETslash and \HT.
 For   SUSY models that yield events with low \ETslash and/or \HT (CMSSM
 with $m_\text{0}>1000$, and for SMS models close to the diagonal), the
 ANN's performance is better compared to  the analyses using large
 \ETslash and \HT selection criteria.

 In the case of the
 CMSSM limits and  for a specific choice of parameter values,
 squark masses below ${\sim}700\GeV$ are excluded at $95\%$ CL; and
 similarly
 gluino masses below ${\sim}700\GeV$ are excluded for the region $m_0<700\GeV$. In the region $1000 < m_0 < 3000\GeV$,
 gluino masses
 below  ${\sim}300 \GeV$ are excluded, while the squark mass in the excluded
 models varies in the range from 1000\GeV to  2500\GeV, depending on the value
 $m_0$. In the case of the SMS limits, for gluino masses below
 $\sim$800\GeV, LSP masses below $\sim$400\GeV are excluded. For
 gluino masses above $\sim$800\GeV, no limits on the mass of LSP can be set.

\section{Conclusions}
 A  search  for supersymmetry in events with two opposite-sign leptons in the final state and
 with the use of an artificial neural network has been presented, using the 2011 dataset collected
 with the CMS  experiment.
 This search is complementary to the ones already published  by the CMS collaboration and yields
comparable exclusion limits for high-\ETslash, high-\HT SUSY models. In addition, the
 significantly relaxed criteria on \ETslash and \HT with
 respect to the previously published analyses
 allows for the study of events not addressed by previous searches,
 and provides an independent and complementary probe of this particularly challenging region
 of phase space.  Agreement is observed  between the expectation from the SM and the data, with no
 significant  excess, which results in limits in the CMSSM
($m_0,m_\text{1/2}$) and SMS
($m_{\PSg},m_\text{LSP}$)
 planes.

\section*{Acknowledgements}

\hyphenation{Bundes-ministerium Forschungs-gemeinschaft Forschungs-zentren}
We congratulate our colleagues in the CERN accelerator departments for the
excellent performance of the LHC and thank the technical and
administrative staffs at CERN and at other CMS institutes for their
contributions to the success of the CMS effort. In addition, we gratefully
acknowledge the computing centres and personnel of the Worldwide LHC
Computing Grid for delivering so effectively the computing infrastructure
essential to our analyses. Finally, we acknowledge the enduring support
for the construction and operation of the LHC and the CMS detector
provided by the following funding agencies: the Austrian Federal Ministry
of Science and Research; the Belgian Fonds de la Recherche Scientifique,
and Fonds voor Wetenschappelijk Onderzoek; the Brazilian Funding Agencies
(CNPq, CAPES, FAPERJ, and FAPESP); the Bulgarian Ministry of Education,
Youth and Science; CERN; the Chinese Academy of Sciences, Ministry of
Science and Technology, and National Natural Science Foundation of China;
the Colombian Funding Agency (COLCIENCIAS); the Croatian Ministry of
Science, Education and Sport; the Research Promotion Foundation, Cyprus;
the Ministry of Education and Research, Recurrent financing contract
SF0690030s09 and European Regional Development Fund, Estonia; the Academy
of Finland, Finnish Ministry of Education and Culture, and Helsinki
Institute of Physics; the Institut National de Physique Nucl\'eaire et de
Physique des Particules~/~CNRS, and Commissariat \`a l'\'Energie Atomique
et aux \'Energies Alternatives~/~CEA, France; the Bundesministerium f\"ur
Bildung und Forschung, Deutsche Forschungsgemeinschaft, and
Helmholtz-Gemeinschaft Deutscher Forschungszentren, Germany; the General
Secretariat for Research and Technology, Greece; the National Scientific
Research Foundation, and National Office for Research and Technology,
Hungary; the Department of Atomic Energy and the Department of Science and
Technology, India; the Institute for Studies in Theoretical Physics and
Mathematics, Iran; the Science Foundation, Ireland; the Istituto Nazionale
di Fisica Nucleare, Italy; the Korean Ministry of Education, Science and
Technology and the World Class University program of NRF, Republic of
Korea; the Lithuanian Academy of Sciences; the Mexican Funding Agencies
(CINVESTAV, CONACYT, SEP, and UASLP-FAI); the Ministry of Science and
Innovation, New Zealand; the Pakistan Atomic Energy Commission; the
Ministry of Science and Higher Education and the National Science Centre,
Poland; the Funda\c{c}\~ao para a Ci\^encia e a Tecnologia, Portugal; JINR
(Armenia, Belarus, Georgia, Ukraine, Uzbekistan); the Ministry of
Education and Science of the Russian Federation, the Federal Agency of
Atomic Energy of the Russian Federation, Russian Academy of Sciences, and
the Russian Foundation for Basic Research; the Ministry of Science and
Technological Development of Serbia; the Secretar\'{\i}a de Estado de
Investigaci\'on, Desarrollo e Innovaci\'on and Programa Consolider-Ingenio
2010, Spain; the Swiss Funding Agencies (ETH Board, ETH Zurich, PSI, SNF,
UniZH, Canton Zurich, and SER); the National Science Council, Taipei; the
Thailand Center of Excellence in Physics, the Institute for the Promotion
of Teaching Science and Technology of Thailand and the National Science
and Technology Development Agency of Thailand; the Scientific and
Technical Research Council of Turkey, and Turkish Atomic Energy Authority;
the Science and Technology Facilities Council, UK; the US Department of
Energy, and the US National Science Foundation.
Individuals have received support from the Marie-Curie programme and the
European Research Council (European Union); the Leventis Foundation; the
A. P. Sloan Foundation; the Alexander von Humboldt Foundation; the Belgian
Federal Science Policy Office; the Fonds pour la Formation \`a la
Recherche dans l'Industrie et dans l'Agriculture (FRIA-Belgium); the
Agentschap voor Innovatie door Wetenschap en Technologie (IWT-Belgium);
the Ministry of Education, Youth and Sports (MEYS) of Czech Republic; the
Council of Science and Industrial Research, India; the Compagnia di San
Paolo (Torino); and the HOMING PLUS programme of Foundation for Polish
Science, cofinanced from European Union, Regional Development Fund.

\bibliography{auto_generated}

\cleardoublepage \appendix\section{The CMS Collaboration \label{app:collab}}\begin{sloppypar}\hyphenpenalty=5000\widowpenalty=500\clubpenalty=5000\textbf{Yerevan Physics Institute,  Yerevan,  Armenia}\\*[0pt]
S.~Chatrchyan, V.~Khachatryan, A.M.~Sirunyan, A.~Tumasyan
\vskip\cmsinstskip
\textbf{Institut f\"{u}r Hochenergiephysik der OeAW,  Wien,  Austria}\\*[0pt]
W.~Adam, E.~Aguilo, T.~Bergauer, M.~Dragicevic, J.~Er\"{o}, C.~Fabjan\cmsAuthorMark{1}, M.~Friedl, R.~Fr\"{u}hwirth\cmsAuthorMark{1}, V.M.~Ghete, N.~H\"{o}rmann, J.~Hrubec, M.~Jeitler\cmsAuthorMark{1}, W.~Kiesenhofer, V.~Kn\"{u}nz, M.~Krammer\cmsAuthorMark{1}, I.~Kr\"{a}tschmer, D.~Liko, I.~Mikulec, M.~Pernicka$^{\textrm{\dag}}$, D.~Rabady\cmsAuthorMark{2}, B.~Rahbaran, C.~Rohringer, H.~Rohringer, R.~Sch\"{o}fbeck, J.~Strauss, A.~Taurok, W.~Waltenberger, C.-E.~Wulz\cmsAuthorMark{1}
% ...
\vskip\cmsinstskip
\dag:~Deceased\\
1:~~Also at Vienna University of Technology, Vienna, Austria\\
2:~~Also at CERN, European Organization for Nuclear Research, Geneva, Switzerland\\
% ...

\end{sloppypar}
\end{document}